%% file: paper.tex
\documentclass[letterpaper,twocolumn,10pt]{article}
\usepackage{usenix2019_v3}

\usepackage{tikz}
\usepackage{amsmath}
\usepackage{graphicx}
\usepackage{filecontents}
\usepackage{xspace}
\usepackage{color, soul, colortbl}
\usepackage{multirow}
\usepackage{subfigure}
\usepackage{tabularx,booktabs}
\usepackage{url}

\newcommand{\comment}[1]{{}}

\begin{document}

\date{}

\title{\Large \bf Analysis of Server Throughput For Managed Big Data Analytics Frameworks.}


\author{
{\rm Emmanouil Anagnostakis}\\
Institute of Computer Science (ICS), Foundation of Research and Technology -- Hellas (FORTH) \\
Computer Science Department, University of Crete, Greece
\and
{\rm Polyvios Pratikakis}\\
Institute of Computer Science (ICS), Foundation of Research and Technology -- Hellas (FORTH) \\
Computer Science Department, University of Crete, Greece
} 

\maketitle
\input{abstract}
\input{intro}
\input{background}

\input{related}

\input{method}
\input{eval}

\input{future_work}

\input{conclusion}

\bibliographystyle{plain}
\bibliography{paper}

\end{document}

%% file: abstract.tex
\begin{abstract}
Managed big data frameworks, such as Apache Spark and Giraph
    demand a large amount of memory per core to process massive volume
    datasets effectively. The memory pressure that arises from the big data processing leads to
    high garbage collection (GC) overhead. Big data analytics frameworks attempt to remove this overhead by offloading objects to storage devices.
    At the same time, infrastructure providers, trying to address the same problem, attribute more memory to increase memory per instance leaving cores underutilized.
        For frameworks, trying to avoid GC through offloading to storage devices leads to high Serialiation/Deserialization (S/D) overhead.
        For infrastructure, the result is that resource usage is decreased.
    These limitations prevent managed big data frameworks from effectively utilizing the CPU thus leading to low server throughput.

    In this thesis, we conduct a methodological analysis of server throughput for
    managed big data analytics frameworks. More specifically, we examine, whether
    reducing GC and S/D can help increase the effective CPU utilization of the server.
    We use a system called TeraHeap (TH) that moves objects from
        the Java managed heap (H1) to a secondary heap over a fast storage
        device (H2) to reduce the GC overhead and eliminate S/D over data. We focus on analyzing the system's
    performance under the co-location of multiple memory-bound
    instances to utilize all available DRAM and study server throughput. Our detailed methodology includes choosing the DRAM budget for each instance and how to distribute
        this budget among H1 and Page Cache (PC). We try two different distributions for the DRAM budget, one with more H1
        and one with more PC to study the needs of both approaches.
    We evaluate both techniques under 3 different memory-per-core scenarios using Spark and Giraph with native JVM or JVM with TeraHeap.
	We do this to check throughput changes when memory capacity increases.

    Our experimental results show that increasing memory per core does not help reach max server throughput for analytics. Effective solutions
    for this problem is using systems like TeraHeap that offload objects from the managed heap without
    increasing the CPU load. Moving large parts of the heap to fast storage, decreases the DRAM GB per core needs and increases the utilization of the server. Finally, we also include a cost
    estimation to show that using an approach like TeraHeap could reduce monetary cost by up
    to 50\% for running big data analytics in a world
    cluster like Amazon's EC2 or Google Cloud Platform or Microsoft
    Azure Cloud, which are available to everyone.
\end{abstract}

%% file: intro.tex
\section{Introduction}
\label{sec:intro}
With the exponential growth of data in various fields such as
healthcare and social media, managed big data frameworks (e.g, Apache
Spark \cite{Spark} and Apache Giraph \cite{Giraph}) require large
amount of DRAM per core for data processing. During the processing, they generate large
amount of objects in the managed heap that span multiple computation
stages. The memory pressure that arises in the managed heap leads to
frequent garbage collection (GC) cycles. Frequent GCs waste CPU cycles 
and prevent application execution.

On the one hand, to reduce the frequency of GC and optimize performance, big data
frameworks offload objects from the managed heap to storage devices. However, these
objects need to be serialized to byte streams to be stored in the storage
device or to be deserialized into memory objects to be loaded back to memory. 
This practice leads to high serialization/deserialization overhead.
On the other hand infrastructure providers, trying to address the same problem increase memory per framework instance that runs in the server. This leaves CPU cores underutilized. 

Co-locating workloads aims to increase available resource utilization
thus increasing the throughput in server. 
In order to maximize throughput, the number of instances
increase to utilize all available DRAM. The result of this
practice is that the underlying machine runs out of memory, while the
 overhead of GC and S/D is still high. The remaining GC and S/D
overheads lead to the problem of wasting the CPU resources
to do unuseful work. This leads to the conclusion that the avalaible memory per core is
not enough for the Garbage Collector, S/D and the application.

The memory per core problem can better be understood when looking at
the resource usage and the characteristics of the servers of big
companies e.g. Alibaba and Facebook. When looking at the results of
Alibaba's traces analyses (\cite{Alibaba}, \cite{Alibaba1},
\cite{Alibabacolocated}) we see that memory usage is at an average of
80\%, while CPU usage stays at 40\%. This trace clearly shows that
DRAM utilization is high, while the CPU is under-utilized. In
Facebook's Twine presentation \cite{Twine}, they used a cluster of
machines where each machine had 40 cores and 80 GB DRAM. This means
that ratio of GB for memory per core was 2. The same ratio is shown in Facebook's Yosemite
\cite{Yosemite}. This shows that memory
capacity for each core is low while DRAM usage is high compared to the CPU usage. Most of the time many the CPU cores are
going to be idle because a few of them will be enough to carry out the
work.

To address the problem of DRAM capacity limitation, recent work
proposed solutions that extend the managed heaps over local flash
storage devices (e.g., NVMe SSD) or remote memory. On the one hand,
TMO \cite{TMO} offloads cold memory to fast storage devices using
a memory scheduling mechanism. On the other hand, CFM \cite{CFM}
utilizes remote DRAM as swap memory in order to increase total memory capacity
and reduce memory pressure. Of both works, only CFM shows
evaluation against managed big data analytics frameworks. However, this evaluation
includes only one Spark workload and is not focused on analytics.

This thesis provides a methodological analysis of server throughput 
focused on managed big data analytics frameworks.
We investigate the off-heap direction of offloading the objects from the
managed heap to fast storage devices.
Specifically, we use TeraHeap (TH) \cite{TeraHeap}, a secondary managed
memory-mapped heap over an NVMe storage device, which is used to hold
the long lived objects instead of the main managed Java Heap. TeraHeap
1) eliminates Serialization/Deserialization overheads posed by this
kind of frameworks when moving data off-heap to/from fast storage
devices 2) reduces GC pauses drastically over the secondary heap. By
using TeraHeap, we aim to investigate the impact of reducing GC and S/D
to server throughput compared to Native Spark and Giraph. We divide all the available DRAM
in our machine to 2,4 and 8 even budgets to run experiments with co-located instances.
We do this to utilize all available DRAM and then check CPU utilization to understand throughput.
First, we run each instance isolated to analyze performance and be able to study the interference when adding more
co-located instances. We run each individual workload with a different Spark or Giraph instance in a cgroup.
We do this, to limit the memory budget for each instance. Memory budget is
the summary of Java Heap, IO Cache (Linux Page Cache) and JVM native memory. We choose
the Java Heap (H1) ratio over the total DRAM budget based on RedHat's decisions
for running containers as a baseline. We also run experiments with more Page Cache (PC) ratio than H1
to investigate Page Cache affection to the performance. We show performance of both Native Spark-Giraph and Spark-Giraph with TH in 3 different
memory per core scenarios, 4 GB per core, which is the current trend and 8 and 16 GB per core 
as possible future trends. We do this to study the changes to server throughput as memory-per-core increases. 
We evaluate both offloading techniques by running 2 widely used
managed big data frameworks, Apache Spark and Giraph. We
specificaly run 4 different workloads with Spark with 4 and 8 GB per core.
We run 2 different workloads with Giraph with 8 and 16 GB per core.
We compare TeraHeap with the native Spark and Giraph distributions under workload
co-location and analyze their performance using several metrics like
GC, S/D, I/O, CPU cycles and CPU utilization. Finally, we estimate the cost of running these
experiments in public world clusters like Amazon EC2, Google Cloud Platform (GCP)
and Microsoft Azure Cloud to see possible benefits of either of the two techniques.

Our experimental results show that increasing memory per core does not guarantee reaching max throughput for managed big data frameworks.
A solution is to move the managed heap over fast storage devices in order to offload objects like TeraHeap and Panthera \cite{Panthera}.
Furthermore, reducing GC and S/D by offloading the heap to fast storage devices improves effective CPU utlization up to 59\% in CPU cycles for Spark and also leaves place to run more co-located instances in the server for both frameworks.
Finally, we also include a cost estimation to show that reducing GC and S/D could reduce monetary spendings by up to 50\% for running big data
analytics, in a world cluster like EC2, GCP or Microsoft Azure Cloud, which are available to
the public.

To summarize, this thesis makes the following contributions: 
\begin{itemize}
    \item{A detailed methodology for running co-located Apache Spark and Giraph
        workloads with or without TeraHeap. 
		We show the interference impact of running multiple co-located managed big data frameworks workloads. We also show that, increasing DRAM capacity is not the solution to the problem of server throughput. First, DRAM density cannot scale further. Therefore, increasing memory-per-core allows more instances to run in the server, but the overheads of GC and S/D remain, because the heap size still is not enough. This leads to the conclusion that these overheads are the obstacle to reach max throughput. Moreover, decreasing GC and S/D, increases the number of co-located instances that can be executed in the server as well.}

    \item{A cost estimation for running our experiments in real-world
        cloud platforms like Amazon EC2, Google Cloud Platform and Microsoft
        Azure. This estimation shows that decreasing GC and S/D leads to less spendings, because money is not wasted to overheads.}
\end{itemize}

%

%% file: background.tex
\section{Background}
\label{sec:background}

In this section, we describe how TeraHeap eliminates GC and S/D.

TeraHeap is a system that eliminates S/D
and GC overheads for a large portion of the data in managed big data
analytics frameworks. TeraHeap extends the Java virtual machine
(JVM) to use a second, high-capacity heap (H2) over a fast storage
device that coexists alongside the regular heap (H1). It eliminates
S/D by providing direct access to objects in H2 and reduces GC
by avoiding costly GC scans over objects in H2. Frameworks use
TeraHeap through its hint-based interface without modifications to
the applications that run on top of them.
TeraHeap provides a hint-based interface that uses key-object opportunism
and enables frameworks to mark objects and indicate when to move
them to H2. During GC, TeraHeap starts from root key-objects and
dynamically identifies the objects to move to H2.

Furthermore, TeraHeap presents a unified heap
with the aggregate capacity of H1 and H2, where scans over H2
during GC are eliminated, to avoid expensive device I/O. To achieve
this, TeraHeap organizes H2 into regions with similar-lifetime objects. 
For space reclamation, the collector reclaims H1
objects as usual. For H2 regions, unlike existing region-based allocators, 
TeraHeap resolves the space-performance trade-off
for reclaiming space differently. Existing allocators reclaim region
space eagerly by moving live objects to another region, which would
generate excessive I/O for storage-backed regions. Instead, 
TeraHeap uses the high capacity of NVMe SSDs to reclaim entire regions
lazily, avoiding slow object compaction on the storage device.

%% file: related.tex
\section{Related Work}

We group the related work in the two following categories:
\begin{itemize}
\item{Works that examine co-location of workloads}
\item{Other analyses on managed big data frameworks}
\end{itemize}

\subsection{Works that examine the co-location of workloads}
To our best knowledge, there is limited work in investigating workload co-location for Managed Big Data Frameworks. Here we refer to some works in this area.

Baig et al. in \cite{NUMA} investigate how Spark-based workloads are impacted by the effects of NUMA-placement decisions. This is something we do not do in our work, because we run our experiments in a single NUMA island to avoid NUMA effects that could complicate the understanding of GC, S/D and the aspects of execution that we investigate. Apart from that difference they investigate the performance of co-located spark workers where each worker runs in a different NUMA island. They count remote memory accesses and context switches in CPU. Chen et al. in \cite{interference} analyze the characteristics of co-located workloads running in containers on the same server from the perspective of hardware events. These events include inctructions per cycle, branch prediction misses and dTLB misses. They also show the execution time of co-located workloads, but they do not provide further analysis or breakdown.

\subsection{Other analyses on managed big data frameworks}

Here we refer to other evaluation works targetting managed big data frameworks.
These works do not provide analyses for workload co-location.

Jiang et al. in \cite{inmem} study the behavior of Spark Workloads in comparison to those of Giraph, CloudSuite, SPEC CPU2006,
TPC-C, and DesktopCloud on system (i.e. disk utilization, memory bandwidth)  and microarchitectural level (instructions per cycle). This work
also provides an analysis for Spark and Giraph examining the behaviour from a different scope than ours. However, it does not provide a breakdown to the execution time of the workloads (i.e. GC, S/D) or CPU utilization analysis.
Ousterhout et al. \cite{makingsense} provide a methodology based on dynamic logging and profiling for quantifying performance bottlenecks in distributed computation frameworks, and use it to analyze the Spark’s performance. They refer and measure S/D, GC and CPU utilization but they don't refer to co-located workloads or target other frameworks. Batarfi et al. \cite{giraphgraphalytics} analyze the performance of many graph processing frameworks including Giraph. They provide results on RAM usage, CPU utilization and execution time. However, they investigate a different aspect from execution time. They break it down to the time taken by each phase of the workload execution.  They also only show results for graph processing and do not target other areas like machine learning as we do. Furthermore, their work is evaluated only aganst Spark.

%% file: method.tex
\section{Experimental Methodology}
\label{sec:method}

In this section we discuss our methodological decisions.

Our methodology answers the following questions:
\begin{itemize}
	\item{What workloads did we choose to run for our experiments and why?}
	\item{How do we investigate the memory per core problem?}
	\item{How do we choose the configurations for running the co-located experiments?}
	\item{Is cost a contributing factor to pursuing higher throughput for a server?}
\end{itemize}

\subsection{Workloads}
For our experiments with Spark, we selected four specific workloads from two
different categories of the Spark Bench suite \cite{Spark-Bench}: Page Rank (PR) and Connected
Component (CC) from GraphX \cite{GraphX} and Linear Regression (LinR) and Logistic Regression (LogR)
from MLLib \cite{MLLib}. For Giraph, we choose PageRank and Community detection 
using label propagation (CDLP) from LDBC Graphalytics \cite{ldbc}. The primary reason for selecting these workloads for Spark is that
they represent different types of algorithms: PR and CC are graph-based workloads, while LinR and LogR are machine learning
workloads. Giraph is a graph processing framework so we only used graph workloads. All of these
workloads are well-established and commonly used for benchmarking big
data analytics systems, making them a suitable choice for our
experiments. Overall, the selection of these workloads allows us to
evaluate the performance of Spark and Giraph in a variety of contexts. Furthermore, it allows us to
provide insights into the performance of both frameworks with or without using TeraHeap.

\subsubsection{PageRank}
PageRank is a widely used graph-based algorithm that measures the
importance of nodes in a network. It has become a popular benchmark
for evaluating the performance of distributed systems, including big
data analytics systems like Apache Spark and Giraph. PageRank is computationally
intensive and requires significant memory and I/O resources, making it
a suitable workload for evaluating performance of managed big data frameworks. Additionally, PageRank is a
common algorithm in real-world applications, such as search engines
and social networks, making it relevant for practical use cases.

\subsubsection{LinearRegression}
LinearRegression is a machine learning algorithm that is used to
predict numerical values based on input data. It is a well-known and
widely used algorithm in machine learning, and is commonly used for
regression analysis in fields such as economics, finance, and
engineering. LinearRegression is computationally intensive and
requires significant memory and I/O resources, making it a suitable
workload for evaluating the performance of managed big data frameworks.

\subsubsection{Logistic Regression}
LogisticRegression is a machine learning algorithm that is used to
model the probability of a binary or categorical outcome based on one
or more independent variables. It is commonly used in predictive
analytics to classify data based on historical data. In Spark-bench,
LogisticRegression is implemented as a machine learning workload,
where the dataset is represented as an RDD of feature vectors and
labels. The LogisticRegression workload involves training a logistic
regression model on the dataset, using an iterative optimization
algorithm such as gradient descent. The workload is computationally
intensive and requires a significant amount of memory to store the
dataset and model parameters, therefore a suitable choise for our experiments..

\subsubsection{Connected Component}
ConnectedComponent is a graph algorithm that is used to identify the
connected components of a graph. It is commonly used in social network
analysis to identify clusters of users with similar interests or
relationships. In Spark-bench, ConnectedComponent is implemented as a
graph processing workload, where the graph is represented as an RDD of
edges and vertices. The ConnectedComponent workload involves iterating
over the graph, identifying the connected components of each node, and
merging the components as necessary. The workload is computationally
intensive and requires a significant amount of memory to store the
graph, therefore a suitable choise for our experiments.

\subsubsection{Community Detection Label Propagation}
The Community Detection using Label Propagation (CDLP) workload, another key component of the Graphalytics benchmark, aims to identify communities within a graph based on label propagation techniques. The CDLP workload assigns labels to nodes iteratively, with each node adopting the most frequently occurring label among its neighbors. This iterative process propagates labels throughout the graph, eventually converging to stable communities. Community detection is a fundamental task in graph analysis, enabling researchers to uncover groups of nodes that exhibit strong internal connectivity. It has applications in social network analysis, recommendation systems, and anomaly detection. The CDLP workload in the Graphalytics benchmark provides a standardized evaluation of graph processing systems' performance in terms of community detection scalability, convergence, and accuracy. By benchmarking CDLP, researchers and practitioners can compare the efficiency and effectiveness of different graph processing platforms and algorithms for community detection tasks.

\subsection{Memory per core}

\begin{table}[thbp]
  \centering
  \caption{Configurations. WL = workload, FW = framework, DS = dataset, Mem.= total memory, M/C = memory per core, Phys. Cores = physical cores}
  \label{tab:setups}
  \begin{tabular}{|c|c|c|c|c|c|c|}
    \hline
	  \textbf{WL} & \textbf{FW} & \textbf{DS (GB)} & \textbf{Mem.} & \textbf{M/C} & \textbf{\#Phys. Cores}\\
    \hline
	  PR & Spark & 8 & 32 & 4 & 8 \\
	  PR & Spark & 8 & 64 & 8 & 8 \\
	  LinR & Spark & 64 & 32 & 4 & 8 \\
	  LinR & Spark & 64 & 64 & 8 & 8 \\
	  LogR & Spark & 64 & 32 & 4 & 8 \\
	  LogR & Spark & 64 & 64 & 8 & 8 \\
	  CC & Spark & 8 & 32 & 4 & 8 \\
	  CC & Spark & 8 & 64 & 8 & 8 \\
	  PR & Giraph & 13 & 64 & 8 & 8 \\
	  CDLP & Giraph & 13 & 64 & 8 & 8 \\
	  PR & Giraph & 13 & 128 & 16 & 8 \\
	  CDLP & Giraph & 13 & 128 & 16 & 8 \\
    \hline
  \end{tabular}
\end{table}

We investigate performance in three memory per core scenarios to check if throughput increases as memory availability increases. Our main focus is 4 GB per core which is the next possible trend in datacenters based on \ref{sec:intro}. The others
are 8 and 16 GB per core which is a possible future trend. For the 4 GB per core scenario, we use 32 GB memory with 8 cores.
In this setup, we run 4 workloads with Native Spark and Spark using TH. Giraph can't run with 4 GB per core with any of the two
configurations. For 8 GB per core, we use 64 GB memory with 8 cores. 
In this setup, we run 4 workloads with Native Spark and Spark using TH and 2 workloads with Native Giraph and Giraph using TH. For the 16 GB memory per core, we use 128 GB of memory with 16 cores. For this setup, we run 2 workloads with Native Giraph and Giraph using TeraHeap. We choose 16 GB per core for Giraph, because it experiences more memory pressure than Spark and it cannot even run with 4 GB DRAM per core. For 8 GB memory per core we are able to run only a few experiments with TH. This happens, because it does not have a very aggresive memory offloading mechanism and cannot offload the heap properly. Table \ref{tab:setups} summarizes all setups with the coresponding workloads and datasets.

\subsection{Choosing the configurations to run the co-located experiments}
To utilize all the available DRAM of the machine, we choose a simple method. 
We divide the total DRAM of the machine by the number of co-located workloads we run for each experiment.
For simplicity and time management, we choose the numbers 1,2,4,8. However, there are 2 problems with that decision
than we need to overcome. The first is that we need a number that is exactly divided by these numbers to give the same amount of DRAM to
all cgroups. The second is that we need to leave memory to the OS for system tasks that are executed along with the system reserved memory of 2 GB. For the 4 GB memory per core, we utilize 24 out of 32 GB total DRAM and leave the rest 8 GB to the OS for system reserved memory and system tasks. For the 8 GB memory per core scenario, we utilize 56 GB out of 64 GB total DRAM and leave the rest 8 GB to the OS. For the 16 GB memory per core scenario we utilize 120 out of 128 GB total DRAM and leave the rest 8 GB to the OS. 
In all scenarios, we choose the number closer to the total DRAM. After having perfomed these calculations, we run each experiment isolated and break down the execution time to know how each experiment performs isolated. Then we run the co-located experiments and study the interference in execution. For the co-located experiments, we run the same workload with the same dataset size for all instances. We do this for simplicity of explaining the results. For all experiments, we disable swap memory.
We use cgroups \cite{cgroups} to restrict the DRAM for all processes in a single instance of Spark and Giraph.
For cgroups, we choose as a baseline an 80\% of total cgroup DRAM budget for H1 as RedHat does for its cgroup containers from June 2023 \cite{redhatblog}.
The rest 20\% remains to the OS to be used as Page Cache.
For TeraHeap, we also run experiments with 40\% for H1 to investigate what happens when Page Cache dominates H1.
In some experiments TeraHeap requires more than 20\% for OS so we make an adjustment to the cgroup budget.
For Native Spark and Giraph, we do not report results for those experiments as we saw that Page Cache adjustments make no difference.

\subsection{Cost estimation}
Renting servers is a common practice for organizations requiring
computational resources, and the question arises as to whether
reducing the monetary cost is possible by achieving higher throughput
and faster workload completion. The relationship between cost
reduction and achieving higher throughput on rented servers is indeed
significant. By optimizing server performance, efficiently utilizing
resources, implementing workload scheduling, and improving
productivity, organizations can realize cost savings. Achieving higher
throughput and faster workload completion can lead to a reduced rental
duration, minimizing the time and associated costs of server usage.
Efficient resource utilization and workload scheduling contribute to
cost reduction by minimizing the number of servers required and
maximizing their utilization. Rental pricing models that take into
account resource utilization or data processed can further reduce
costs for organizations achieving higher throughput. Additionally,
improved productivity resulting from higher throughput and faster
workload completion enhances overall efficiency, allowing
organizations to accomplish more work within the same rental period
and reducing rental expenses. Therefore, pursuing higher throughput
and faster workload completion offers tangible benefits in terms of
monetary cost reduction for organizations renting servers. 

We estimate the cost of our experiments in real-world public clusters, to show that increasing throughput by
decreasing GC and S/D leads to avoiding wasting money in overheads when renting servers. We
chose a variety of providers like Amazon \cite{EC2}, Google \cite{GCP} and Microsoft \cite{Azure}. This
way, we covered the most known providers and platforms someone would
choose to run their workloads on. We chose 3 machines from each
platform identical to the specifications of our 32, 64 and 128 GB DRAM
machines. These are the cheapest machines of that particular category
offered by the platform. We then used the platform's pricing
calculator to estimate the cost of renting that machine for the time
needed for each configuration to finish execution of all instances. We noticed, that the price for
renting the storage device is really amenable to the cost for renting the machine.

%% file: eval.tex
\section{Evaluation}
\label{sec:eval}

In this section we report and analyze our experiments and we also state our conclusions.

\subsection{Native Spark Configuration}
We use Spark v3.3.0 (\cite{Building}, \cite{Tuning}, \cite{Conf}, \cite{Monitoring}) with Kryo Serializer \cite{Kryo}, a state-of-the-art highly optimized S/D Library for Java that Spark recommends. We run Spark
with Native OpenJDK8 \cite{JDK8} as a baseline. We use the Parallel Scavenge
garbage collector which is the one TeraHeap is implemented for.
Parallel Scavenge is also the go-to collector for applications that
need high throughput like Spark. We use one executor with 8
threads for each instance of Spark we deploy on our server \cite{TeraHeap}. Spark storage level
is configured to MEMORY-AND-DISK to place executor memory (heap) in DRAM and cache RDDs \cite{RDD}
in the on-heap cache, up to 50\% of the total heap size. Any remaining
RDDs are serialized in the off-heap cache over an NVMe SSD. This
device is also used by Spark for shuffling. 

\subsection{Native Giraph Configuration}
We run Giraph with Native OpenJDK8 \cite{JDK8} as a baseline. We use the Parallel Scavenge
garbage collector. We use one executor with 8 threads for each instance of Giraph we deploy on our server \cite{TeraHeap}. Native Giraph offloads messages and edges to the storage device.

\subsection{Spark-Giraph configurations for TeraHeap}
\subsubsection{Spark Configuration}
The configuration for TeraHeap is pretty much the same as for Native
Spark, with some necessary differences. TeraHeap
is mapped to a different storage device (NVMe) than that Spark is
using for shuffling. We do this in order for TeraHeap to utilize its
device to its fullest. MMIO allows TeraHeap Spark to run in
MEMORY-ONLY storage level as Spark remains unaware of using any device and
the OS takes control of the I/O.

\subsubsection{Giraph Configuration}
For Giraph, we map TeraHeap to a different NVMe storage device that the one we
use for Zookeeper. TeraHeap works in the same way as in Spark,
thus Giraph is unaware of the presence of a second heap.

\subsection{Experiments with single instance}

In this section, we run the single instance experiments and provide an explanation of their performance to use it later to study the interference between single and co-located experiments.
These experiments map one to one to the co-located experiments of the next section.
DRAM per core is added to the figure titles to show relation between this mapping, and not because it has any impact
for single instance performance. For all figures, each configuration is described with memory capacity for H1 + memory for OS in GB and a label that denotes the division of memory e.g. N2 is 1/2 of total DRAM for Native, N4 is 1/4 of total DRAM for Native. TH H1 denotes 80\% memory for H1 and TH PC denotes 40\% memory for H1 to investigate the PC scenario. LinR and LogR experiments with 10 GB DRAM for H1 and 4 for OS for TH that do not match the 80\% budget baseline are conducted this way, because the OS needs 1 extra GB for cache. This is not an Out of memory (OOM) error for H1 but an adjustment to memory budget.
X axis shows each configuration. Y axis shows execution time in seconds. All missing configurations in the figure are OOM experiments.

\begin{figure}[thbp]
\centering
    \includegraphics[width=\linewidth]{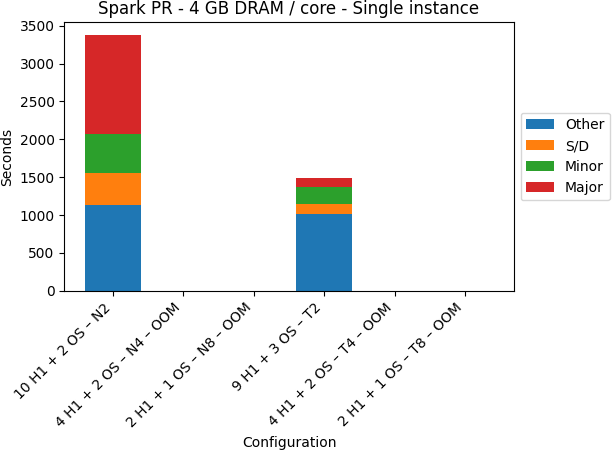}
    \caption{ Execution time breakdown for single instances of Spark
        Page Rank for the 4 GB memory-per-core scenario.}
    \label{fig:pr32_single}
        \includegraphics[width=\linewidth]{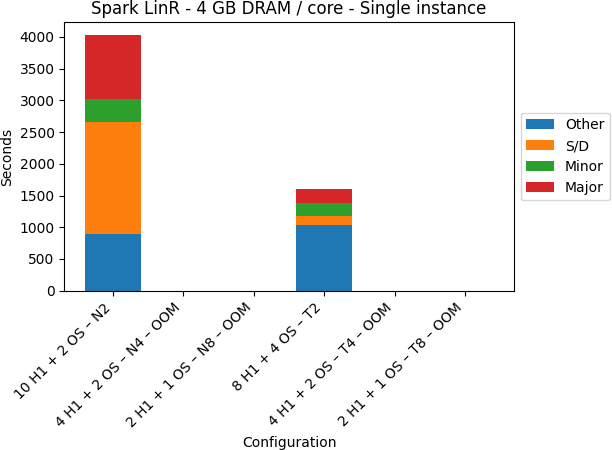}
    \caption{Execution time breakdown for single instances of Spark
        Linear Regression for the 4 GB memory-per-core scenario.}
    \label{fig:linr32_single}
\end{figure}

\begin{figure}[thbp]
        \centering
    \includegraphics[width=\linewidth]{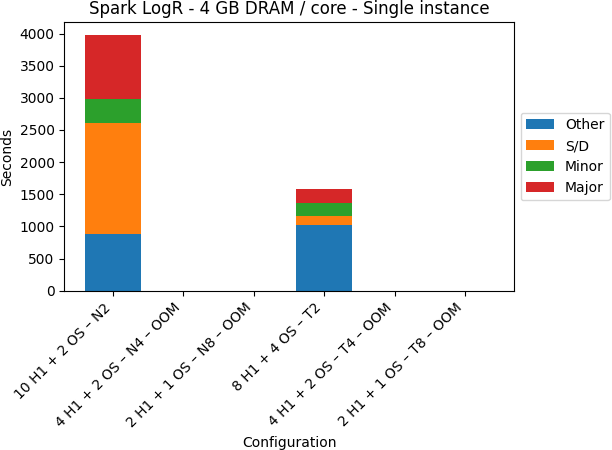}
    \caption{Execution time breakdown for single instances of Spark
    Logistic Regression for the 4 GB memory-per-core scenario.}
    \label{fig:logr32_single}

    \includegraphics[width=\linewidth]{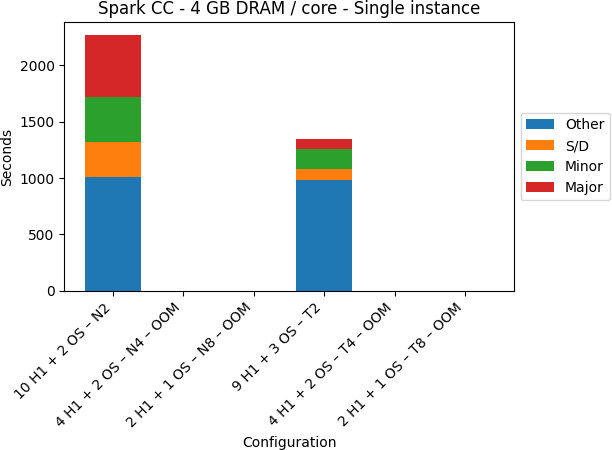}
    \caption{Execution time breakdown for single instances of Spark
	Connected Component for the 4 GB memory-per-core scenario.}
    \label{fig:cc32_single}
\end{figure}

\begin{figure}[thbp]
\centering
    \includegraphics[width=\linewidth]{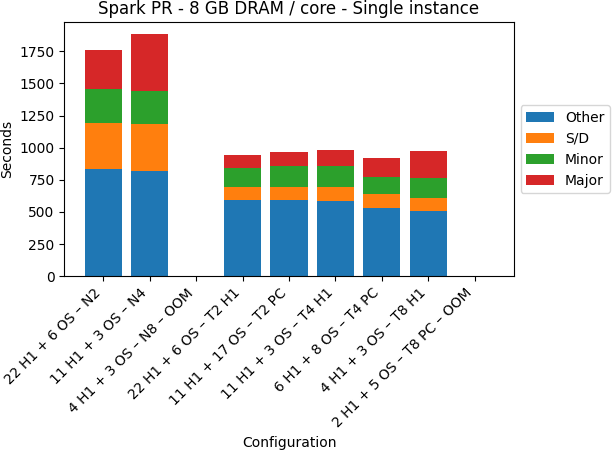}
    \caption{ Execution time breakdown for single instances of Spark
	Page Rank for the 8 GB memory-per-core scenario.}
    \label{fig:pr64_single}
	\includegraphics[width=\linewidth]{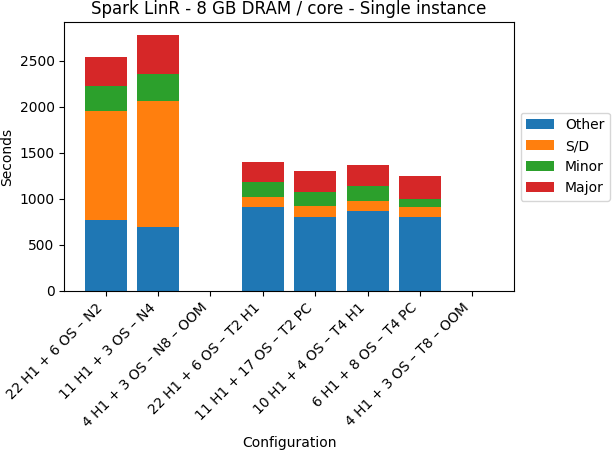}
    \caption{Execution time breakdown for single instances of Spark
	Linear Regression for the 8 GB memory-per-core scenario.}
    \label{fig:linr64_single}
\end{figure}

\begin{figure}[thbp]
        \centering
    \includegraphics[width=\linewidth]{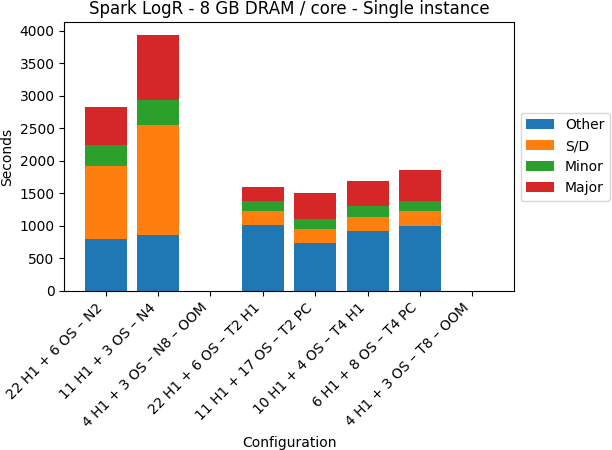}
    \caption{Execution time breakdown for single instances of Spark
    Logistic Regression for the 8 GB memory-per-core scenario.}
    \label{fig:logr64_single}

    \includegraphics[width=\linewidth]{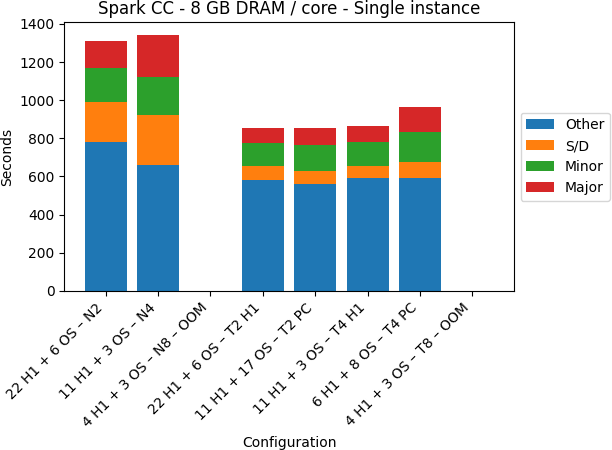}
    \caption{Execution time breakdown for single instances of Spark
    Connected Component for the 8 GB memory-per-core scenario.}
    \label{fig:cc64_single}
\end{figure}

\begin{figure}[thbp]
        \centering
    \includegraphics[width=\linewidth]{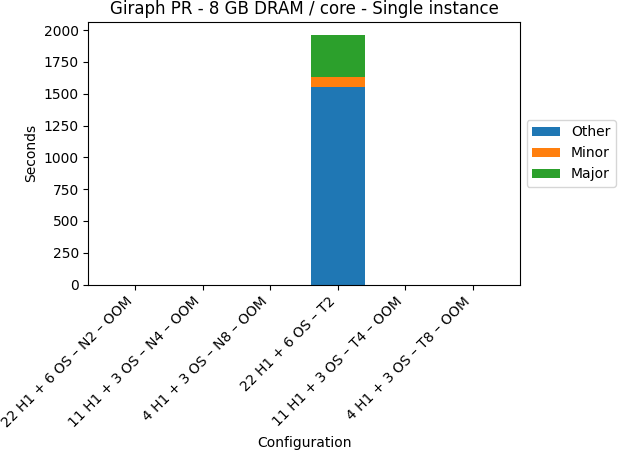}
    \caption{Execution time breakdown for single instances of Giraph
	Page Rank for the 8 GB memory-per-core scenario.}
    \label{fig:g_pr64_single}
    \includegraphics[width=\linewidth]{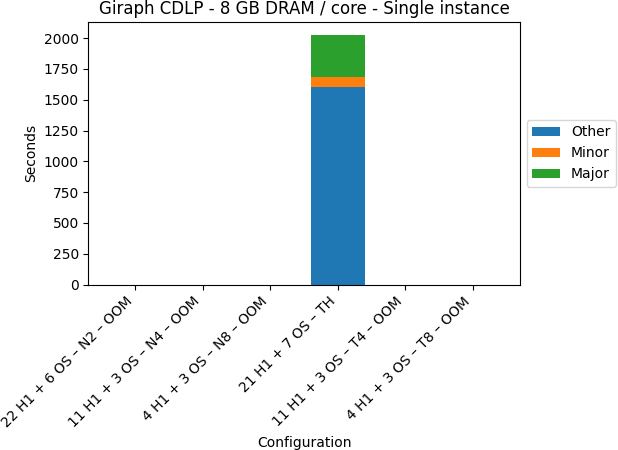}
    \caption{Execution time breakdown for single instances of Giraph
    Community Detection Label Propagation for the 8 GB memory-per-core scenario.}
    \label{fig:g_cdlp64_single}
\end{figure}

\begin{figure}[thbp]
        \centering
    \includegraphics[width=\linewidth]{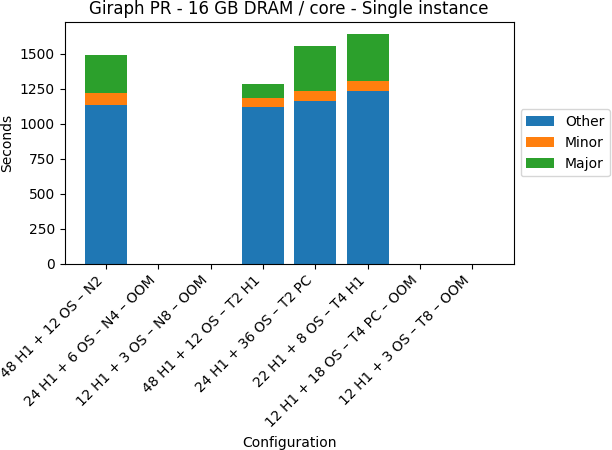}
    \caption{Execution time breakdown for single instances of Giraph
    Page Rank for the 16 GB memory-per-core scenario.}
    \label{fig:g_pr128_single}
    \includegraphics[width=\linewidth]{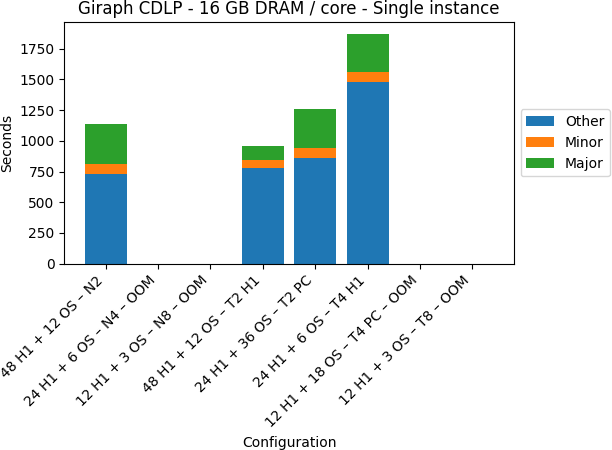}
    \caption{Execution time breakdown for single instances of Giraph
    Community Detection Label Propagation for the 16 GB memory-per-core scenario.}
    \label{fig:g_cdlp128_single}
\end{figure}

\begin{figure}[thbp]
\centering
    \includegraphics[width=\linewidth]{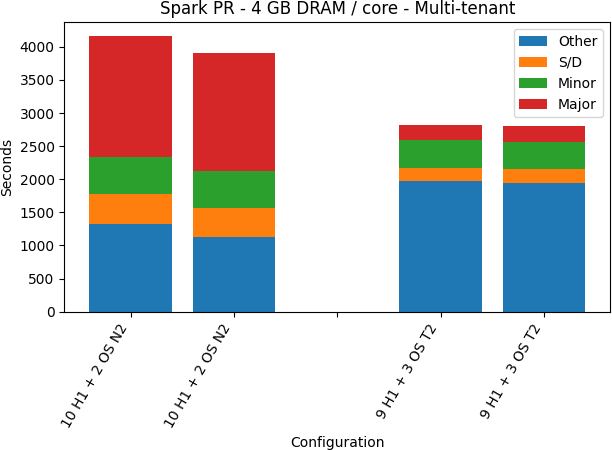}
    \caption{Execution time breakdown for co-located instances of Spark
    Page Rank in the 4 GB memory-per-core scenario.}
    \label{fig:pr32}
        \includegraphics[width=\linewidth]{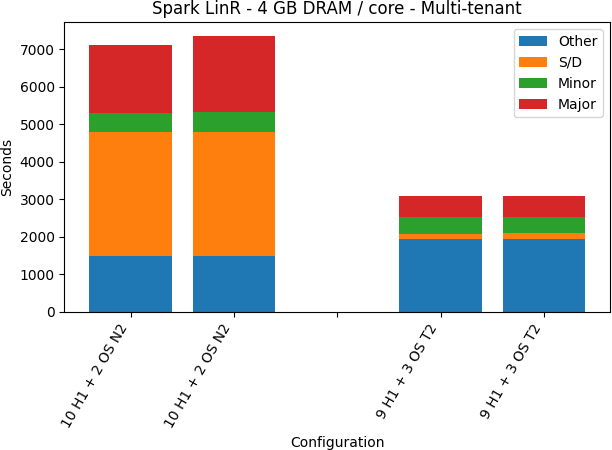}
    \caption{Execution time breakdown for co-located instances of Spark
    Linear Regression in the 4 GB memory-per-core scenario.}
    \label{fig:linr32}
\end{figure}

Figure \ref{fig:pr32_single} shows single instance performance with Page Rank for Native and TH Spark. These experiments correspond to the co-located runs of figure \ref{fig:pr32}. The first bar shows performance of Native Spark for 12 GB DRAM. The second bar shows execution breakdown of TH Spark for 12 GB DRAM. This figure shows that Native Spark suffers from GC, while TH absorbs this overhead.

Figure \ref{fig:linr32_single} shows single instance performance with Linear Regression for Native and TH Spark. These experiments correspond to the co-located runs of figure \ref{fig:linr32}. The first bar shows performance of Native Spark for 12 GB DRAM. The second bar shows execution breakdown of TH Spark for 12 GB DRAM. This figure shows that Native Spark suffers from GC and S/D, while TH absorbs these overheads.

\begin{figure}[thbp]
        \centering
    \includegraphics[width=\linewidth]{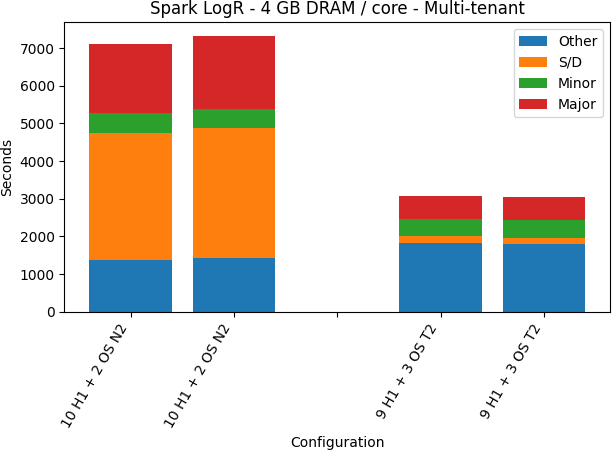}
    \caption{Execution time breakdown for co-located instances of Spark
    Logistic Regression in the 4 GB memory-per-core scenario.}
    \label{fig:logr32}

    \includegraphics[width=\linewidth]{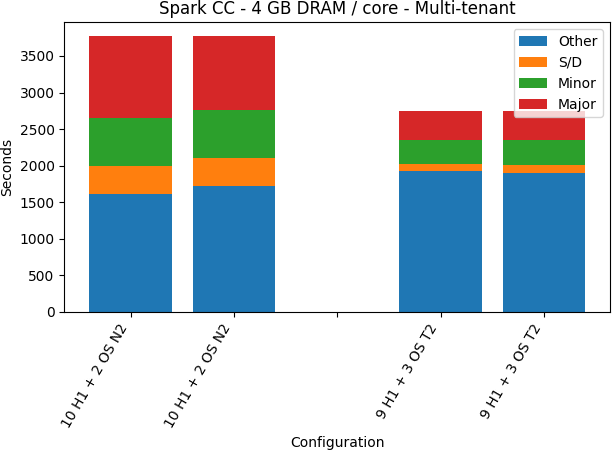}
    \caption{Execution time breakdown for co-located instances of Spark
    Connected Component in the 4 GB memory-per-core scenario.}
    \label{fig:cc32}
\end{figure}

Figure \ref{fig:logr32_single} shows single instance performance with Logistic Regression for Native and TH Spark. These experiments correspond to the co-located runs of figure \ref{fig:logr32}. The first bar shows performance of Native Spark for 12 GB DRAM. The second bar shows execution breakdown of TH Spark for 12 GB DRAM. This figure shows that Native Spark suffers from GC and S/D, while TH absorbs these overheads.

Figure \ref{fig:cc32_single} shows single instance performance with Connected Component for Native and TH Spark. These experiments correspond to the co-located runs of figure \ref{fig:cc32}. The first bar shows performance of Native Spark for 12 GB DRAM. The second bar shows execution breakdown of TH Spark for 12 GB DRAM. This figure shows that Native Spark suffers from GC, while TH absorbs this overhead.

\begin{figure}[thbp]
\centering
    \includegraphics[width=\linewidth]{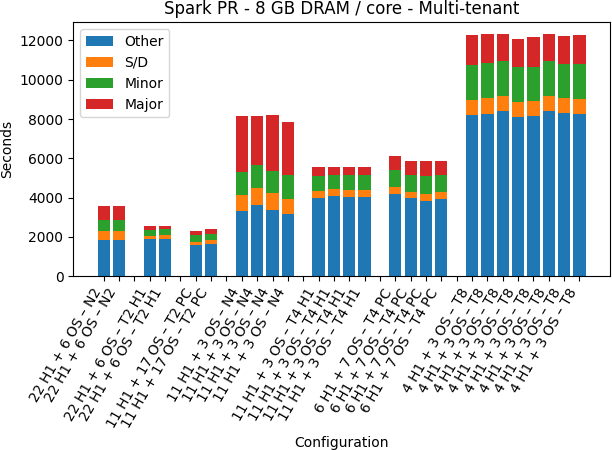}
    \caption{Execution time breakdown for co-located instances of Spark
    Page Rank in the 8 GB memory-per-core scenario.}
    \label{fig:pr64}
	\includegraphics[width=\linewidth]{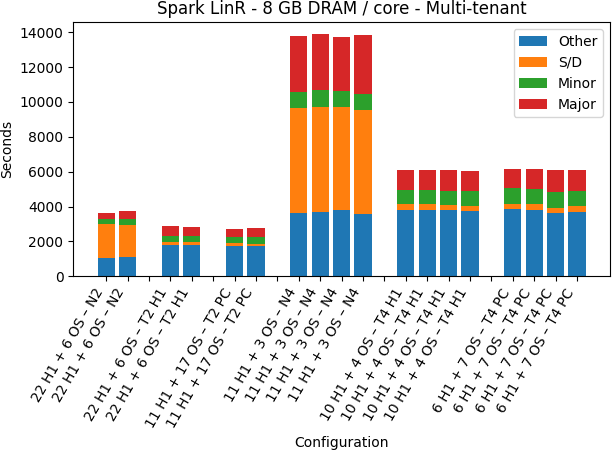}
    \caption{Execution time breakdown for co-located instances of Spark
    Linear Regression in the 8 GB memory-per-core scenario.}
    \label{fig:linr64}
\end{figure}

Figure \ref{fig:pr64_single} shows single instance performance with Page Rank for Native and TH Spark. These experiments correspond to the co-located runs of figure \ref{fig:pr64}. The first two bars show performance of Native Spark for 28 and 14 GB DRAM. When H1 decreases, Native suffers from longer and more frequent GC cycles, thus we see an increment to Major GC. S/D and other time remain the same as Read/Write traffic remains the same. The rest four bars show performance for TH Spark for 28 (80\% and 40\% for H1), 14 (80\% and 40\% for H1) and 7 (80\% for H1) GB DRAM. For TH PC there is no memory for the system. As we said in our methodology, for TeraHeap we investigate setups with DRAM budgets where both H1 and PC dominate. As H1 decreases for TeraHeap, we see an increase to Major GC in the last 2 bars. Other time and S/D remain the same.

Figure \ref{fig:linr64_single} shows single instance performance with Linear Regression for Native and TH Spark. These experiments correspond to the co-located runs of figure \ref{fig:linr64}. The first two bars show performance of Native Spark for 28 and 14 GB DRAM. When H1 decreases, Native suffers from longer and more frequent GC cycles thus we see an increment to Major GC. S/D has a slight increase because of increased read traffic caused by memory pressure. Write traffic remains the same because objects in Spark are immutable. The rest four bars show performance for TH Spark for 28 (80\% and 40\% for H1) and 14 (71\% and 40\% for H1) GB DRAM. As H1 decreases for TeraHeap, we see an increase to Major GC in the last 2 bars. Other time shows slight differences because of cache size. That can be seen from the second and third bar which have the same amount for H1 and a big difference in cache. S/D remains the same.

\begin{figure}[thbp]
        \centering
    \includegraphics[width=\linewidth]{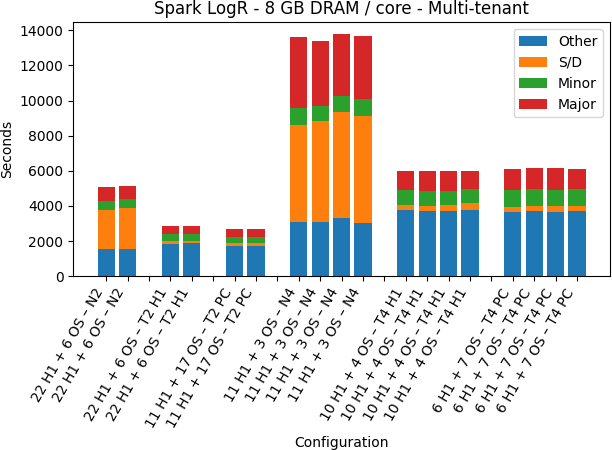}
    \caption{Execution time breakdown for co-located instances of Spark
    Logistic Regression in the 8 GB memory-per-core scenario.}
    \label{fig:logr64}

    \includegraphics[width=\linewidth]{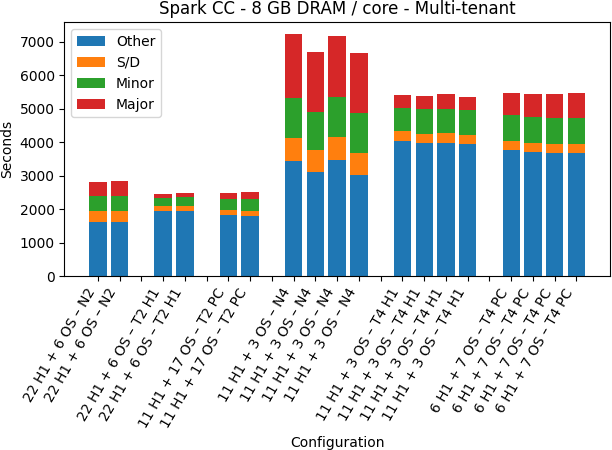}
    \caption{Execution time breakdown for co-located instances of Spark
    Connected Component in the 8 GB memory-per-core scenario.}
    \label{fig:cc64}
\end{figure}

Figure \ref{fig:logr64_single} shows single instance performance with Logistic Regression for Native and TH Spark. These experiments correspond to the co-located runs of figure \ref{fig:logr64}. The first two bars show performance of Native Spark for 28 and 14 GB DRAM. When H1 decreases Native, suffers from longer and more frequent GC cycles, thus we see a significant increment to Major GC. S/D has a huge increase of almost 30\% because of increased read traffic caused by memory pressure. Write traffic remains the same because objects in Spark are immutable. The rest four bars show performance for TH Spark for 28 (80\% and 40\% for H1) and 14 (71\% and 40\% for H1) GB DRAM. As H1 decreases for TeraHeap, we see some notable differences to GC. Other time shows differences because of cache size. That can be seen from the second and third bar which have the same amount for H1 and a big difference in cache. S/D remains the same.

Figure \ref{fig:cc64_single} shows single instance performance with Connected Component for Native and TH Spark. These experiments correspond to the co-located runs of figure \ref{fig:cc64}. The first two bars show performance of Native Spark for 28 and 14 GB DRAM. When H1 decreases, Native suffers from longer and more frequent GC cycles, thus we see an increment to Major GC. S/D remains the same. Write traffic remains the same because objects in Spark are immutable. The rest four bars show performance for TH Spark for 28 (80\% and 40\% for H1) and 14 (80\% and 40\% for H1) GB DRAM. As H1 decreases for TeraHeap, we see an increase to Minor GC in the last bar. Other time and S/D remain the same.

Giraph cannot run at all with 4 GB memory per core.

Figures \ref{fig:g_pr64_single} and \ref{fig:g_cdlp64_single} show performance only for TH with 80\% budget for H1, because all other experiments are OOM, thus  we cannot provide a comparison with other experiments.

\begin{figure}[thbp]
        \centering
    \includegraphics[width=\linewidth]{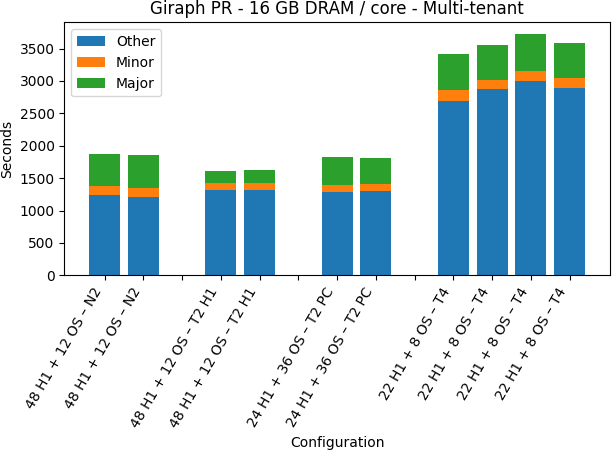}
    \caption{Execution time breakdown for co-located instances of Giraph
    Page Rank in the 16 GB memory-per-core scenario.}
    \label{fig:g_pr128}
    \includegraphics[width=\linewidth]{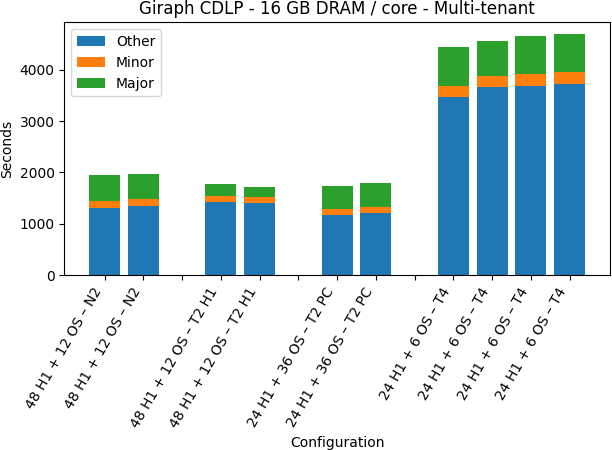}
    \caption{Execution time breakdown for co-located instances of Giraph
    Community Detection Label Propagation in the 16 GB memory-per-core scenario.}
    \label{fig:g_cdlp128}
\end{figure}

Figure \ref{fig:g_pr128_single} shows single instance performance with Page Rank for Native and TH Giraph. These experiments correspond to the co-located runs of figure \ref{fig:g_pr128}. The first bar shows performance of Native Giraph for 60 GB DRAM. The rest three bars show performance for TH Giraph for 60 (80\% and 40\% for H1) and 30 (80\% for H1) GB DRAM. As H1 decreases for TeraHeap, we see an increase to Major GC  and Other time. Other time changes by both H1 and Page Cache differences. We see that H1 affects writes in a significant way, because objects are mutable in Giraph and decreasing H1 creates more traffic to TeraHeap. Page Cache mostly affects read traffic. These can be seen from the progression of the bars in other time.

Figure \ref{fig:g_cdlp128_single} shows single instance performance with Community Detection Label Propagation for Native and TH Giraph. These experiments correspond to the co-located runs of figure \ref{fig:g_pr128}. The first bar shows performance of Native Giraph for 60 GB DRAM. The rest three bars show performance for TH Giraph for 60 (80\% and 40\% for H1) and 30 (80\% for H1) GB DRAM. As H1 decreases for TeraHeap, we see an increase to Major GC  and Other time. Other time changes by both H1 and Page Cache differences. We see that H1 affects writes in a significant way, because objects are mutable in Giraph, and decreasing H1 creates more traffic to TeraHeap. Page Cache mostly affects read traffic. These can be seen from the progression of the bars in other time.

In all Spark experiments we see that, H1 has significant impact for Native, while for TeraHeapi, H1 is significant too, but not as significant as for Native. For Native, we saw no differences with variable Page Cache sizes for any of the experiments, thus we do not show them here. For TH, PC shows improvements of 5\% to 7\% for ML workloads, except the LinR experiment that maps to the co-located experiment for 8 GB per core. Number of GCs and Read/Write traffic figures are not included because all preserve the same pattern described above. For Native Spark the number of GCs and read traffic increases significantly as H1 decreases. For TH Spark, number of GCs also increase slightly as H1 decreases and read/write traffic remains the same. Read traffic increases slightly as PC decreases for TH. For Giraph, H1 also affects read/write traffic significantly for both Native and TH and PC decreases read traffic significantly for TH.

\subsection{Experiments with co-located instances}

Here, we look at the co-located experiments of Spark and Giraph in all memory per core categories.
We run these experiments to see whether increasing memory-per-core helps increasing server throughput by reducing GC and S/D for frameworks
and increasing number of instances for infrastructure.
Any runs that are not shown should be considered experiments that run Out of memory (OOM) for H1.
We do not include them in the figures, because they are exactly the same configurations that run OOM in their corresponding single instance run. These can be seen in the figures of the previous subsection.
N2 (T for TeraHeap) means that we have a co-located experiment with 2 instances of Native Spark or Giraph.
Average throughout is the result of the division of the result of the multiplication of the number of instances with dataset size (same per instance) and the execution time of the slowest instance in execution. Realizations on other time are included in a different subsection.
All results are rounded to the upper bound integer except costs, because for monetary cost even small amounts are significant.
X axis shows each configuration. Y axis shows execution time in seconds.

\begin{figure}[thbp]
        \centering
    \includegraphics[width=\linewidth]{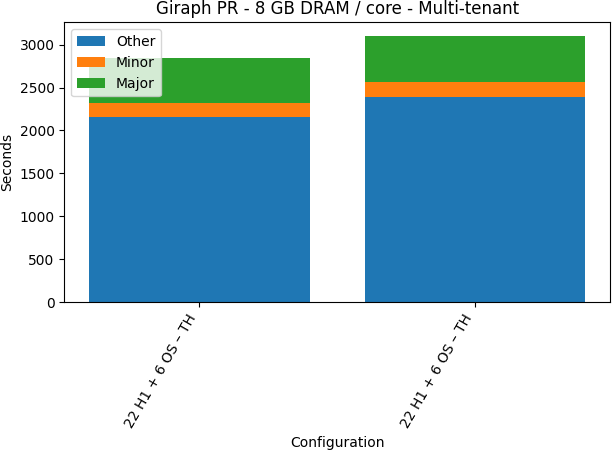}
    \caption{Execution time breakdown for co-located instances of Giraph
    Page Rank in the 8 GB memory-per-core scenario.}
    \label{fig:g_pr64}
    \includegraphics[width=\linewidth]{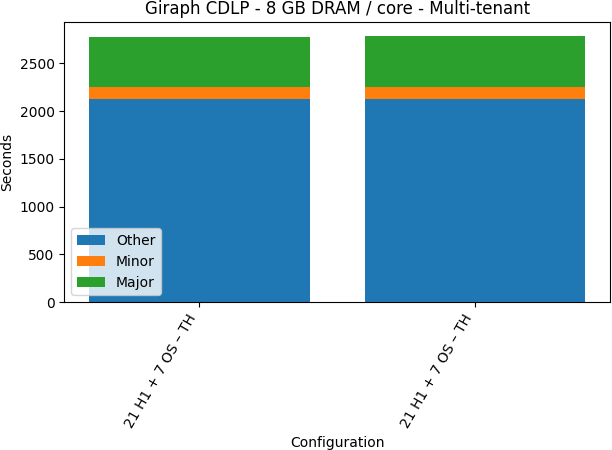}
    \caption{Execution time breakdown for co-located instances of Giraph
    Community Detection Label Propagation in the 8 GB memory-per-core scenario.}
    \label{fig:g_cdlp64}
\end{figure}

\begin{figure}[thbp]
        \centering
        \includegraphics[width=\linewidth]{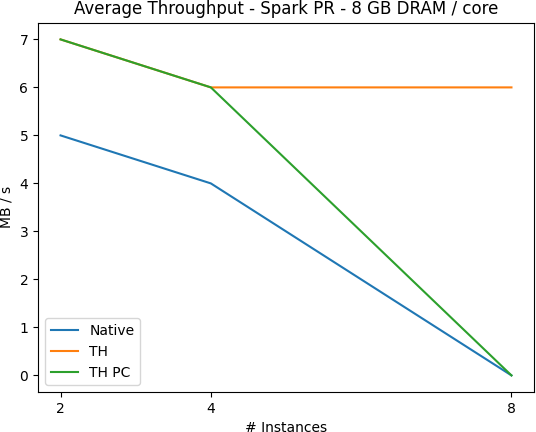}
    \caption{Native and TeraHeap Spark average throughput
        as the number of instances increases under 8 GB DRAM per core running Page Rank.}
\label{fig:pr_64_thr}
        \includegraphics[width=\linewidth]{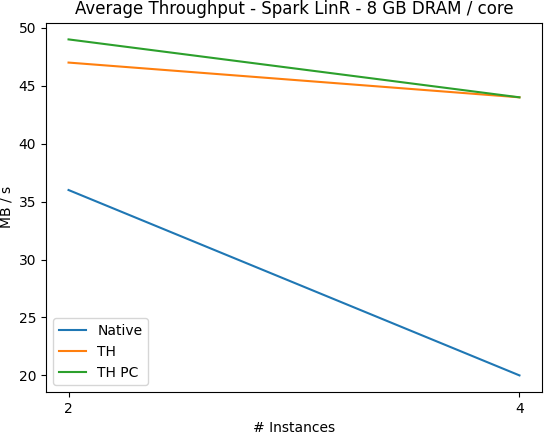}
    \caption{Native and TeraHeap Spark average throughput
        as the number of instances increases under 8 GB DRAM per core running Linear Regression.}
                \label{fig:linr_64_thr}
\end{figure}

\begin{figure}[thbp]
        \centering
        \includegraphics[width=\linewidth]{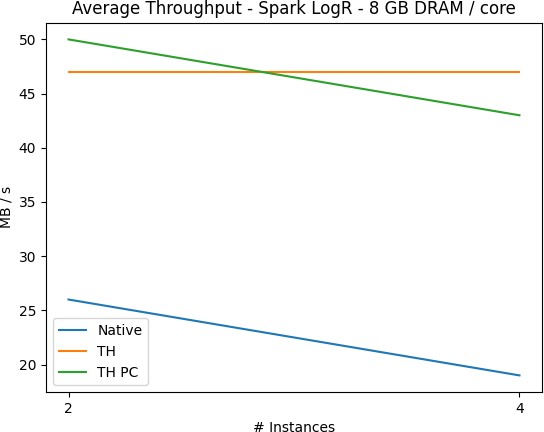}
    \caption{Native and TeraHeap Spark average throughput
        as the number of instances increases under 8 GB DRAM per core running Logistic Regression.}
                \label{fig:logr_64_thr}
        \includegraphics[width=\linewidth]{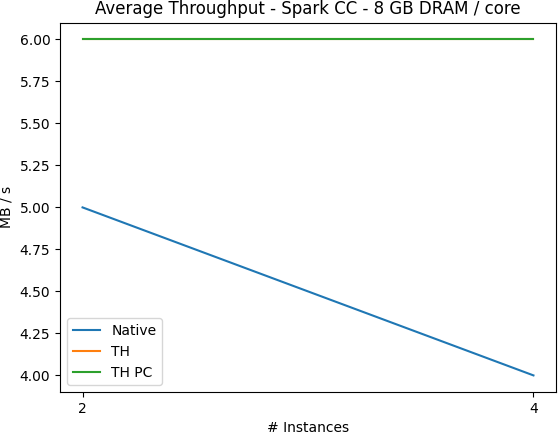}
    \caption{Native and TeraHeap Spark average throughput
        as the number of instances increases under 8 GB DRAM per core running Connected Component.}
                \label{fig:cc_64_thr}
\end{figure}

\begin{figure}[thbp]
        \centering
        \includegraphics[width=\linewidth]{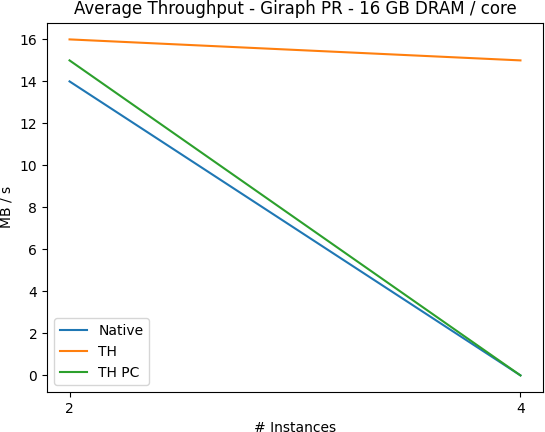}
    \caption{Native and TeraHeap Giraph average throughput
        as the number of instances increases under 16 GB DRAM per core running Page Rank.}
        \label{fig:g_pr128_thr}
        \includegraphics[width=\linewidth]{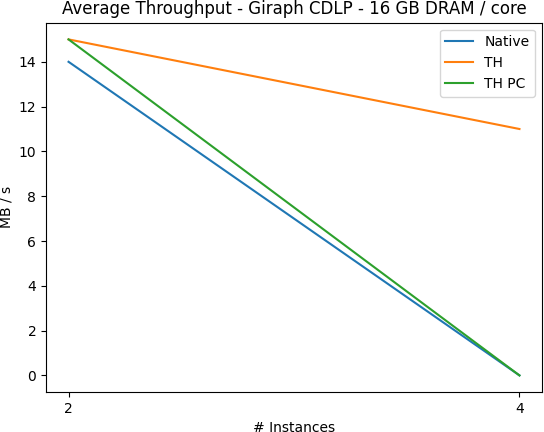}
    \caption{Native and TeraHeap Giraph average throughput
        as the number of instances increases under 16 GB DRAM per core running Page Rank.}
        \label{fig:g_cdlp128_thr}
\end{figure}

We explain each figure from 4 aspects:
\begin{itemize}
\item{The differences in the time breakdown while number of instances increase for each configuration.}
\item{A comparison between the different configurations while instances increase.}
\item{Interference between the single instance and co-located instances}
\item{A comparison between H1 and Page Cache dominating configurations}
\item{Realizations on performance difference between different memory per core scenarios}
\end{itemize}

\subsubsection{4 GB DRAM per core}

Figure \ref{fig:pr32} shows execution time of co-located
Native-TeraHeap Spark instances running PageRank with 8 GB
dataset per instance in the 4 GB DRAM per core scenario.
In the graph, we witness the performance of 2 runs. The first run is with 2 co-located Native Spark instances.
The other run is with 2 co-located TH Spark instances with H1 dominating Page Cache.
We could run the experiment where PC dominates H1, but we did not, because of lack of time. 
Each instance of the Native run uses 10 GB DRAM for H1 (Java Heap) and 2 GB for rest of the services.
The TH run uses 9 GB DRAM for H1 and 3 GB for Page Cache for each instance.

Considering the first aspect, we do not have the needed runs to analyze it.

From the second aspect, we see that as Native Spark starves from more GC and S/D, TeraHeap nearly eliminates these overheads. TeraHeap has 32\% speedup and 33\% more average throughput for 2 instances when compared to the corresponding Native runs.

Figure \ref{fig:linr32} shows execution time of co-located
Native-TeraHeap Spark instances running LinearRegression with 8 GB
dataset per instance in the 4 GB DRAM per core scenario.
In the graph, we witness the performance of 2 runs. The first run is with 2 co-located Native Spark instances.
The other run is with 2 co-located TH Spark instances with H1 dominating Page Cache.
We could run the experiment where PC dominates H1, but we did not, because of lack of time.
Each instance of the Native run uses 10 GB DRAM for H1 (Java Heap) and 2 GB for rest of the services.
The TH run uses 8 GB DRAM for H1 and 4 GB for Page Cache for each instance.

Considering the first aspect, we do not have the needed runs to analyze it.

From the second aspect, we see that as Native Spark starves from more GC and S/D, TeraHeap nearly eliminates these overheads. TeraHeap has 58\% speedup and 59\% more average throughput for 2 instances when compared to the corresponding Native runs.

Figure \ref{fig:logr32} shows execution time of co-located
Native-TeraHeap Spark instances running PageRank with 8 GB
dataset per instance in the 4 GB DRAM per core scenario.
In the graph, we witness the performance of 2 runs. The first run is with 2 co-located Native Spark instances.
The other run is with 2 co-located TH Spark instances with H1 dominating Page Cache.
We could run the experiment where PC dominates H1, but we did not, because of lack of time.
Each instance of the Native run uses 10 GB DRAM for H1 (Java Heap) and 2 GB for rest of the services.
The TH run uses 8 GB DRAM for H1 and 4 GB for Page Cache for each instance.

Considering the first aspect, we do not have the needed runs to analyze it.

From the second aspect, we see that as Native Spark starves from more GC and S/D, TeraHeap nearly eliminates these overheads. TeraHeap has 58\% speedup and 59\% more average throughput for 2 instances when compared to the corresponding Native runs.

Figure \ref{fig:cc32} shows execution time of co-located
Native-TeraHeap Spark instances running PageRank with 8 GB
dataset per instance in the 4 GB DRAM per core scenario.
In the graph, we witness the performance of 2 runs. The first run is with 2 co-located Native Spark instances.
The other run is with 2 co-located TH Spark instances with H1 dominating Page Cache.
We could run the experiment where PC dominates H1, but we did not, because of lack of time.
Each instance of the Native run uses 10 GB DRAM for H1 (Java Heap) and 2 GB for rest of the services.
The TH run uses 9 GB DRAM for H1 and 3 GB for Page Cache for each instance.

Considering the first aspect, we do not have the needed runs to analyze it.

From the second aspect, we see that as Native Spark starves from more GC and S/D, TeraHeap nearly eliminates these overheads. TeraHeap has 32\% speedup and 33\% more average throughput for 2 instances when compared to the corresponding Native runs.

\subsubsection{8 GB DRAM per core}

Figure \ref{fig:pr64} and \ref{fig:pr_64_thr} show execution time and average throughput of co-located
Native-TeraHeap Spark instances running PageRank with 8 GB
dataset per instance in the 8 GB DRAM per core scenario.
Starting from the left of the graph, the first 6 bars show the
performance of 3 runs. The first run is with 2 co-located Native Spark instances.
Another run with 2 co-located TH Spark instances with H1 dominating Page Cache,
and a third run with 2 co-located TH Spark instances where Page Cache dominates H1.
Each instance of the first 2 runs uses 22 GB DRAM for H1 (Java Heap) and 6 GB for rest of the services.
The third run uses 11 GB DRAM for H1 and 17 GB for Page Cache for each instance. 
The next 12 bars show the performance of another 3 runs. The first run is with 4 co-located Native Spark instances.
Another run with 4 co-located TH Spark instances with H1 dominating Page Cache,
and a third run with 4 co-located TH Spark instances where Page Cache dominates H1.
Each instance of the first run uses 11 GB DRAM for H1 (Java Heap) and 3 GB for rest of the services.
The second run uses 11 GB DRAM for H1 and 3 GB for Page Cache for each instance.
The third run uses 6 GB DRAM for H1 and 8 GB for Page Cache for each instance.
The last 8 bars refer to 8 co-located instances of TeraHeap Spark only. 
We were unable to decrease H1 enough to run 8 co-located instances of Native Spark,
because JVM runs out of memory. Each instance of the run uses 4 GB DRAM for H1 (Java Heap) and 3 GB for Page Cache.

Considering the first aspect, we see that Minor and Major GC increase dramatically for Native Spark along with significant increase to Other time. Minor and Major GC differences are witnessed, because the heap capacity decreases and that causes memory pressure. TeraHeap Spark shows a slight increase to Major GC, while the number of instances increases. This is because of the decreasing heap capacity. We suspect device throughput reaching its limit with increasing number of instances, as the cause to other time for both Native and TH. S/D is completely absorbed by MMIO. For Native Spark 2 co-located instances have 55\% speedup in execution time compared to 4 co-located instances, and provide 20\% more average throughput. For TH H1 2 co-located instances have 40\% speedup in execution time compared to 4 co-located instances and provide 14\% more average throughput. For TH 8 co-located instances have 50 and 83\% speedup against 4 and 2 instances accordingly.

From the second aspect, as instances increase in the server the benefit gap between Native and TeraHeap Spark becomes bigger. As Native Spark starves from more GC and S/D, TeraHeap maintains its benefits. TeraHeap has 50 and 25\% speedup for 2 and 4 instances when compared to the corresponding Native runs. If we compare TeraHeap 8 instances to the 4 instances of Native TeraHeap has 33\% worse performance but 33\% more average throughput.

Figure \ref{fig:linr64} and \ref{fig:linr_64_thr} show the execution time and average throughput of co-located
Native-TeraHeap Spark instances running LinearRegression with 64 GB
dataset per instance in the 8 GB DRAM per core scenario.
Starting from the left of the graph, the first 6 bars show the
performance of 3 runs. The first run is with 2 co-located Native Spark instances.
Another run with 2 co-located TH Spark instances with H1 dominating Page Cache,
and a third run with 2 co-located TH Spark instances where Page Cache dominates H1.
Each instance of the first 2 runs uses 22 GB DRAM for H1 (Java Heap) and 6 GB for rest of the services.
The third run uses 11 GB DRAM for H1 and 17 GB for Page Cache for each instance.
The rest 12 bars show the performance of another 3 runs. The first run is with 4 co-located Native Spark instances.
Another run with 4 co-located TH Spark instances with H1 dominating Page Cache
and a third run with 4 co-located TH Spark instances where Page Cache dominates H1.
Each instance of the first run uses 11 GB DRAM for H1 (Java Heap) and 3 GB for rest of the services.
The second run uses 10 GB DRAM for H1 and 4 GB for Page Cache for each instance.
The third run uses 6 GB DRAM for H1 and 8 GB for Page Cache for each instance.

Considering the first aspect, we see that GC and S/D increase dramatically for Native Spark along with significant increase to Other time. GC differences are witnessed because the heap capacity decreases, and that causes memory pressure. TeraHeap Spark shows a slight increase to Major GC while the number of instances increases. This is because of the decreased heap capacity. We suspect device throughput reaching its limit with increasing number of instances as the cause to other time for both Native and TH. S/D is completely absorbed by MMIO. For Native Spark 2 co-located instances have 71\% speedup in execution time compared to 4 co-located instances and provide 46\% more average throughput. For TH H1 2 co-located instances have 50\% speedup in execution time compared to 4 co-located instances, and provide 8\% more average throughput. For TH PC performance is the same with TH H1.

From the second aspect, as instances increase in the server the benefit gap between Native and TeraHeap Spark becomes bigger. As Native Spark starves from more GC and S/D, TeraHeap maintains its benefits. That is shown by the speedups where TeraHeap has 25\% and 57\% speedup and 48\% and 66\% more average throughput for 2 and 4 instances when compared to the corresponding Native runs.

Figures \ref{fig:logr64} and \ref{fig:logr_64_thr} show execution time and average throughput of co-located
Native-TeraHeap Spark instances running Logistic Regression with 64 GB
dataset per instance in the 8 GB DRAM per core scenario.
Starting from the left of the graph, the first 6 bars show the
performance of 3 runs. The first run is with 2 co-located Native Spark instances.
Another run with 2 co-located TH Spark instances with H1 dominating Page Cache
and a third run with 2 co-located TH Spark instances where Page Cache dominates H1.
Each instance of the first 2 runs uses 22 GB DRAM for H1 (Java Heap) and 6 GB for rest of the services.
The third run uses 11 GB DRAM for H1 and 17 GB for Page Cache for each instance. 
The next 12 bars show the performance of another 3 runs. The first run is with 4 co-located Native Spark instances.
Another run with 4 co-located TH Spark instances with H1 dominating Page Cache
and a third run with 4 co-located TH Spark instances where Page Cache dominates H1.
Each instance of the first run uses 11 GB DRAM for H1 (Java Heap) and 3 GB for rest of the services.
The second run uses 10 GB DRAM for H1 and 3 GB for Page Cache for each instance.
The third run uses 6 GB DRAM for H1 and 7 GB for Page Cache for each instance.

Considering the first aspect, we see that GC and S/D increase dramatically for Native Spark along with significant increase to Other time. GC differences are witnessed because the heap capacity decreases, and that causes memory pressure. TeraHeap Spark shows a slight increase to Major GC while the number of instances increases. This is because of the decreased heap capacity. We suspect device throughput reaching its limit with increasing number of instances, as the cause to other time for both Native and TH. S/D is completely absorbed by MMIO. For Native Spark 2 co-located instances have 62\% speedup in execution time compared to 4 co-located instances and provide 27\% more average throughput. For TH H1 2 co-located instances have 50\% speedup in execution time compared to 4 co-located instances and provides the same throughput. For TH PC performance is the same with TH H1. 

From the second aspect, as instances increase in the server, the benefit gap between Native and TeraHeap Spark becomes bigger. As Native Spark starves from more GC and S/D, TeraHeap maintains its benefits. TeraHeap has 57 and 40\% speedup and 48\% and 66\% increased average throughput for 2 and 4 instances when compared to the corresponding Native runs.

Figure \ref{fig:cc64} and \ref{fig:cc_64_thr} show execution time and average throughput of co-located
Native-TeraHeap Spark instances running Connected Component with 8 GB
dataset per instance in the 8 GB DRAM per core scenario.
Starting from the left of the graph, the first 6 bars show the
performance of 3 runs. The first run is with 2 co-located Native Spark instances.
Another run with 2 co-located TH Spark instances with H1 dominating Page Cache
and a third run with 2 co-located TH Spark instances where Page Cache dominates H1.
Each instance of the first 2 runs uses 22 GB DRAM for H1 (Java Heap) and 6 GB for rest of the services.
The third run uses 11 GB DRAM for H1 and 17 GB for Page Cache for each instance. 
The next 12 bars show the performance of another 3 runs. The first run is with 4 co-located Native Spark instances.
Another run with 4 co-located TH Spark instances with H1 dominating Page Cache
and a third run with 4 co-located TH Spark instances where Page Cache dominates H1.
Each instance of the first run uses 11 GB DRAM for H1 (Java Heap) and 3 GB for rest of the services.
The second run uses 11 GB DRAM for H1 and 3 GB for Page Cache for each instance.
The third run uses 6 GB DRAM for H1 and 8 GB for Page Cache for each instance.

Considering the first aspect, we see that Minor and Major GC increase dramatically for Native Spark along with significant increase to Other time. Minor and Major GC differences are witnessed because the heap capacity decreases, and that causes memory pressure. TeraHeap Spark shows a slight increase to Major GC while the number of instances increases. This is because of the decreasing heap capacity. We suspect device throughput reaching its limit with increasing number of instances, as the cause to other time for both Native and TH. S/D is completely absorbed by MMIO. For Native Spark 2 co-located instances have 57\% speedup in execution time compared to 4 co-located instances and provide 27\% more average throughput. For TH H1 2 co-located instances have 54\% speedup in execution time compared to 4 co-located instances and provides 8\% less throughput.

From the second aspect, as instances increase in the server, the benefit gap between Native and TeraHeap Spark becomes bigger. As Native Spark starves from more GC and S/D, TeraHeap maintains its benefits. TeraHeap has 21 and 15\% speedup and 10\% more throughput for 2 and 4 instances when compared to the corresponding Native runs.

\subsubsection{16 GB DRAM per core}

Figures \ref{fig:g_pr64} and \ref{fig:g_cdlp64} show execution time only for TH Giraph with 80\% budget for H1, because all other experiments are
OOM thus we cannot provide a comparison with other experiments.

Figure \ref{fig:g_pr128} and \ref{fig:g_pr128_thr} show execution time and average throughput of co-located
Native-TeraHeap Giraph instances running Page Rank with 13 GB
dataset per instance in the 16 GB DRAM per core scenario.
Starting from the left of the graph, the first 6 bars show the
performance of 3 runs. The first run is with 2 co-located Native Giraph instances.
Another run with 2 co-located TH Giraph instances with H1 dominating Page Cache,
and a third run with 2 co-located TH Instances instances where Page Cache dominates H1.
Each instance of the first 2 runs uses 48 GB DRAM for H1 (Java Heap) and 12 GB for rest of the services.
The third run uses 24 GB DRAM for H1 and 36 GB for Page Cache for each instance.
The rest 4 bars show the performance of another run. The run is with 4 co-located TeraHeap Giraph instances.
Each instance uses 24 GB DRAM for H1 (Java Heap) and 6 GB for rest of the services.

Considering the first aspect Native Giraph does not scale to 4 instances and runs out of memory. TeraHeap Giraph shows significant increase to Major GC while the number of instances increases. This is because of the decreased heap capacity. We suspect device throughput reaching its limit with increasing number of instances, as the cause to other time. For TH H1, 2 co-located instances have 57\% speedup in execution time, compared to 4 co-located instances, and provide the same average throughput. For TH PC, 2 co-located instances have 51\% speedup in execution time compated to 4 co-located instances, and provide the same average throughput.

From the second aspect, TeraHeap is able to scale to 4 instances, while Native runs out of memory. TeraHeap has 11\% speedup and 13\% more average throughput for 2 instances, when compared to the corresponding Native runs.

Figure \ref{fig:g_cdlp128} and \ref{fig:g_cdlp128_thr} show execution time and average throughput of multiple
Native-TeraHeap Giraph instances running CDLP with 13 GB
dataset per instance in the 16 GB DRAM per core scenario.
Starting from the left of the graph, the first 6 bars show the
performance of 3 runs. The first run is with 2 co-located Native Giraph instances.
Another run with 2 co-located TH Giraph instances with H1 dominating Page Cache,
and a third run with 2 co-located TH Instances instances where Page Cache dominates H1.
Each instance of the first 2 runs uses 48 GB DRAM for H1 (Java Heap) and 12 GB for rest of the services.
The third run uses 24 GB DRAM for H1 and 36 GB for Page Cache for each instance.
The rest 4 bars show the performance of another run. The run is with 4 co-located TeraHeap Giraph instances.
Each instance uses 24 GB DRAM for H1 (Java Heap) and 6 GB for rest of the services.

Considering the first aspect, Native Giraph does not scale to 4 instances and runs out of memory. TeraHeap Giraph shows significant increase to Major GC, while the number of instances increases. This is because of the decreased heap capacity. We suspect device throughput reaching its limit with increasing number of instances, as the cause to other time. For TH H1, 2 co-located instances have 63\% speedup in execution time, compared to 4 co-located instances and provide 27\% more average throughput. For TH PC, 2 co-located instances have 61\% speedup in execution time, compared to 4 co-located instances and 27\% more average throughput. From the second aspect, TeraHeap is able to scale to 4 instances while Native runs out of memory. TeraHeap has 9\% speedup and 7\% more average throughput for 2 instances, when compared to the corresponding Native runs.

\subsubsection{Realizations for other time}
For both Spark and Giraph, we suspect device throughput reaching its limit with increasing number of instances, as the cause to other time for both Native and TH. TH has increased other time compared to Native, because of the IO granularity of entire pages despite Native having increased read traffic to TH. Native knows exactly what objects to read doing small reads while TeraHeap brings unuseful objects to memory. For Giraph, TeraHeap has increased read/write traffic, compared to Native and both the difference in IO methods, and read/write traffic leads to increased other time.

\subsubsection{Realizations on performance difference between different memory per core scenarios}
For Spark we see that 4 GB memory per core is a bound to run more than 2 instances. For Giraph, we see than Native is unable to run any experiments under 4 and 8 GB memory per core, while TH is able to run with 2 instances proving that lacking enough memory per instance is a bound for execution, while avoiding GC and S/D enables execution.

\subsubsection{Interference with single instance}

\begin{table}[thbp]
  \centering
  \caption{Interference for each configuration with co-located instances with corresponding single instance experiment.
	FW = framework, Conf. = configuration, M/C = Memory per core, \#I = Number of instances, Interf. = interference }
  \label{tab:interference}
  \begin{tabular}{|c|c|c|c|c|}
    \hline
	  \textbf{FW} & \textbf{Conf.} & \textbf{M/C (GB)} & \textbf{\#I} & \textbf{Interf. \%} \\
    \hline
          Spark & PR Native & 4 & 2 & 19 \\
          Spark & PR TH & 4 & 2 & 47 \\
	  Spark & PR TH H1 & 8 & 2 & 63 \\
	  Spark & PR TH PC & 8 & 2 & 59 \\
	  Spark & PR TH H1 & 8 & 4 &  82 \\
	  Spark & PR TH PC & 8 & 4 & 84 \\
	  Spark & PR TH & 8 & 8 & 92 \\
          Spark & LINR Native & 4 & 2 & 45 \\
          Spark & LINR TH & 4 & 2 & 48 \\
	  Spark & LINR Native & 8 & 2 & 32  \\
	  Spark & LINR Native & 8 & 4 & 80 \\
	  Spark & LINR TH H1 & 8 & 2 & 52 \\
	  Spark & LINR TH PC & 8 & 2 & 53 \\
	  Spark & LINR TH H1 & 8 & 4 & 78 \\
	  Spark & LINR TH PC & 8 & 4 & 80 \\
	  Spark & LINR Native & 8 & 2 & 49 \\
          Spark & LOGR Native & 4 & 2 & 46 \\
          Spark & LOGR TH & 4 & 2 & 48 \\
	  Spark & LOGR Native & 8 & 2 & 45 \\
	  Spark & LOGR Native & 8 & 4 & 71 \\
	  Spark & LOGR TH H1 & 8 & 2 & 44 \\
	  Spark & LOGR TH PC & 8 & 2 & 44 \\
	  Spark & LOGR TH H1 & 8 & 4 & 73 \\
	  Spark & LOGR TH PC & 8 & 4 & 75 \\
          Spark & CC Native & 4 & 2 & 40 \\
          Spark & CC TH & 4 & 2 & 51 \\
          Spark & CC Native & 8 & 2 & 56 \\
          Spark & CC Native & 8 & 4 & 75 \\
          Spark & CC TH H1 & 8 & 2 & 66 \\
          Spark & CC TH PC & 8 & 2 & 66 \\
          Spark & CC TH H1 & 8 & 4 & 84  \\
          Spark & CC TH PC & 8 & 4 & 76 \\
	  Giraph & PR TH & 8 & 2 & 37 \\
	  Giraph & CDLP TH & 8 & 2 & 27 \\
          Giraph & PR Native & 16 & 2 & 19 \\
          Giraph & PR TH H1 & 16 & 2 & 21 \\
          Giraph & PR TH PC & 16 & 2 & 38 \\
          Giraph & PR TH & 16 & 4 & 55 \\
          Giraph & CDLP Native & 16 & 2 & 41 \\
          Giraph & CDLP TH H1 & 16 & 2 & 45 \\
          Giraph & CDLP TH PC & 16 & 2 & 30 \\
          Giraph & CDLP TH & 16 & 4 & 67 \\
    \hline
  \end{tabular}
\end{table}

Table \ref{tab:interference} shows the percentage of interference i.e. speedup of single instance against the corresponding co-located experiment. For Native Spark for 2 to 4 co-located instances experiments there is 19 to 80\% interference. For TeraHeap Spark for 2 to 4 co-located instances experiments there is 32 to 84\% interference. Both offloading techniques have similar interference ranges which are more than 50\% in half of the experiments. For Native Giraph there is 19\% interference for PR and 41\% for CDLP with 2 co-located instances. The first is really reduced compared to the Native Spark 2 co-located instances experiments. For TH Giraph there is 21 to 67 \% interference. For 4 co-located instances experiments TH Giraph has significantly less interference than Spark. In conclusion we wee that interference increases as number of instances increases for both Spark and Giraph. Experiments with 2 co-located instances and an interference under 50\% have better average throughput than single instance and the same happens for experiments with 4 co-located instances with interference under 25\%. The latter never happens.

\subsubsection{Does H1 or PageCache offer better performance?}
We don't investigate Page Cache-dominated cgroup budgets for Native Spark or Giraph, since we have seen that it does not make a difference. For TeraHeap Spark, Page Cache provides slightly better average throughput for 2 co-located instances in ML. In speedup, this is 5\% for LinR and 6\% for LogR, while for 4 instances H1 dominates PC. For the Spark GraphX experiments, we witness the same average throughput for both 2 and 4 co-located instances experiments. For TH Giraph, H1 dominates PC in terms of average throughput. That is, because H1 affects Write traffic in Giraph and Page Cache absorbs mostly reads. In conclusion, based on average throughput, it seems someone would still choose H1 dominated setups for TeraHeap as well.

\subsubsection{Accuracy of experiments}

\begin{table}[thbp]
  \centering
  \caption{Standard deviation for each configuration and number of co-located instances.
        FW=framework, Conf. = configuration, M/C = memory per core, \#I=number of instances, St. dev.=standard deviation}
  \label{tab:std-dev}
  \begin{tabular}{|c|c|c|c|c|}
    \hline
          \textbf{FW} & \textbf{Conf.} & \textbf{M/C (GB)} & \textbf{\#I} & \textbf{St. dev. \%} \\
    \hline
          Spark & PR Native & 8 & 2 & 2\\
          Spark & PR Native & 8 & 4 & 6\\
          Spark & PR TH H1 & 8 & 2 & 1 \\
          Spark & PR TH H1 & 8 & 4 & 1 \\
          Spark & LINR Native & 8 & 2 & 2 \\
          Spark & LINR Native & 8 & 4 & 3 \\
          Spark & LINR TH H1 & 8 & 2 & 1 \\
          Spark & LINR TH H1 & 8 & 4 & 2 \\
          Spark & LOGR Native & 8 & 2 & 10 \\
          Spark & LOGR Native & 8 & 4 & 0 \\
          Spark & LOGR TH H1 & 8 & 2 & 3 \\
          Spark & LOGR TH H1 & 8 & 4 & 5 \\
          Spark & CC Native & 8 & 2 & 2 \\
          Spark & CC Native & 8 & 4 & 7 \\
          Spark & CC TH H1 & 8 & 2 & 3 \\
          Spark & CC TH H1 & 8 & 4 & 0 \\
          Giraph & PR TH H1 & 8 & 2 & 6 \\
          Giraph & CDLP Native & 16 & 2 & 4 \\
          Giraph & CDLP TH H1 & 16 & 2 & 5 \\
    \hline
  \end{tabular}
\end{table}

We repeated all experiments for 8 and 16 memory per core with 2 and 4 instances for Spark except with TH PC and 2 instances for Giraph a second time to estimate standard deviation. We left these experiments out because of lack of time. Table \ref{tab:std-dev} shows that 
all experiments have less than 7\% standard deviation except one experiment with Spark for 10\%. Also co-located experiments have under 7\% difference in-between the end of execution of each co-located instance except Native Spark CC with 4 co-located instances under 4 GB DRAM per core with 14\%. This is important, because when one instance has finished the interference decreases for the rest.

\subsection{Is the CPU utilization of the application increasing by reducing GC and S/D?}

\begin{figure}[thbp]
        \centering
        \includegraphics[width=\linewidth]{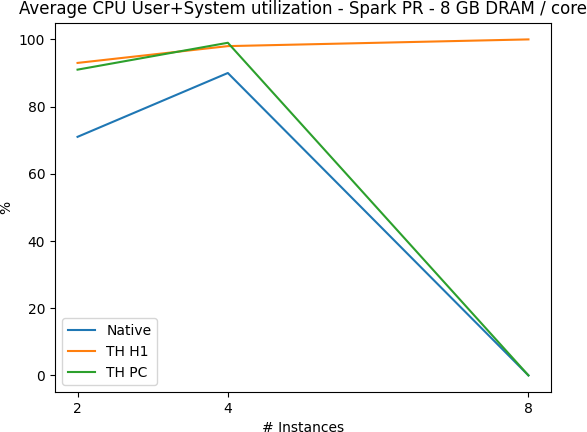}
    \caption{Native and TeraHeap Spark total CPU utilization
        as the number of instances increases under 8 GB DRAM per core running Page Rank.}
\label{fig:pr_64_util}
        \includegraphics[width=\linewidth]{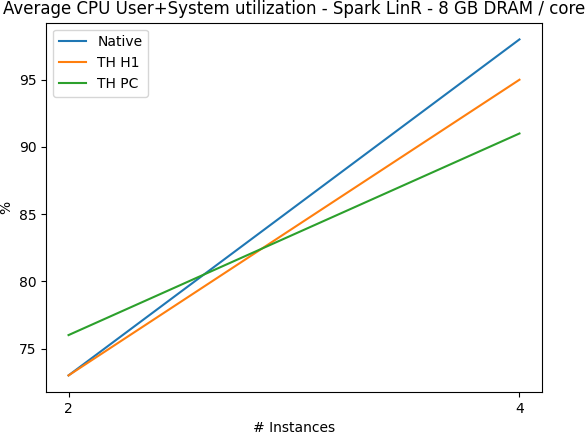}
    \caption{Native and TeraHeap Spark total CPU utilization
        as the number of instances increases under 8 GB DRAM per core running Linear Regression.}
                \label{fig:linr_64_util}
\end{figure}

\begin{figure}[thbp]
        \centering
        \includegraphics[width=\linewidth]{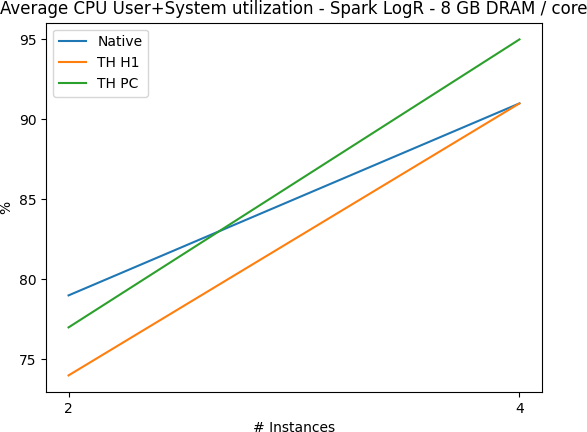}
    \caption{Native and TeraHeap Spark total CPU utilization
        as the number of instances increases under 8 GB DRAM per core running Logistic Regression.}
                \label{fig:logr_64_util}

        \includegraphics[width=\linewidth]{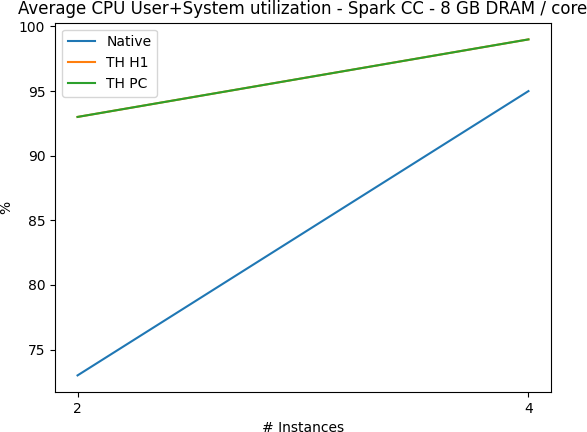}
    \caption{Native and TeraHeap Spark total CPU utilization
        as the number of instances increases under 8 GB DRAM per core running Connected Component.}
                \label{fig:cc_64_util}
\end{figure}

\begin{figure}[thbp]
        \centering
        \includegraphics[width=\linewidth]{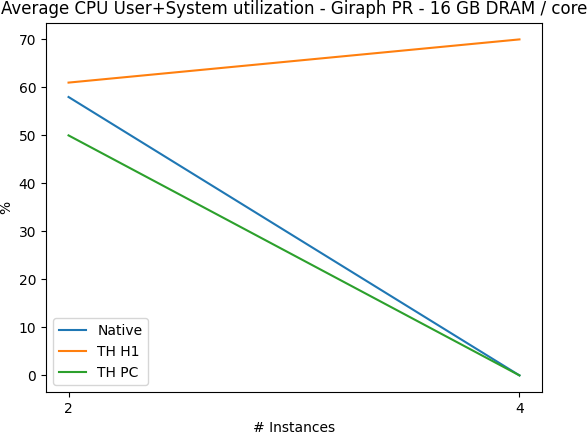}
    \caption{Native and TeraHeap Giraph total CPU utilization
        as the number of instances increases under 8 GB DRAM per core running Page Rank.}
                \label{fig:g_pr_128_util}

        \includegraphics[width=\linewidth]{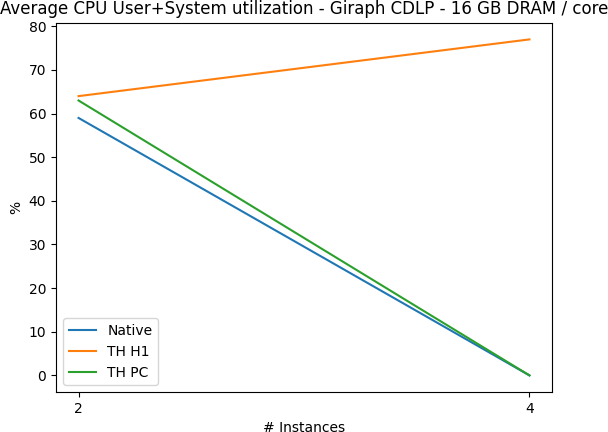}
    \caption{Native and TeraHeap Giraph total CPU utilization
        as the number of instances increases under 8 GB DRAM per core running Community Detection Label Propagation.}
                \label{fig:g_cdlp_128_util}
\end{figure}

\begin{figure}[thbp]
        \centering
        \includegraphics[width=\linewidth]{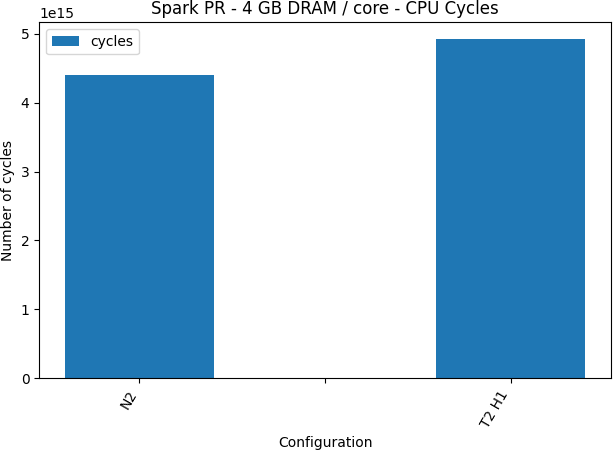}
    \caption{Native and TeraHeap Spark CPU cycles under 4 GB DRAM per core running Page Rank.}
\label{fig:pr32_cycles}
        \includegraphics[width=\linewidth]{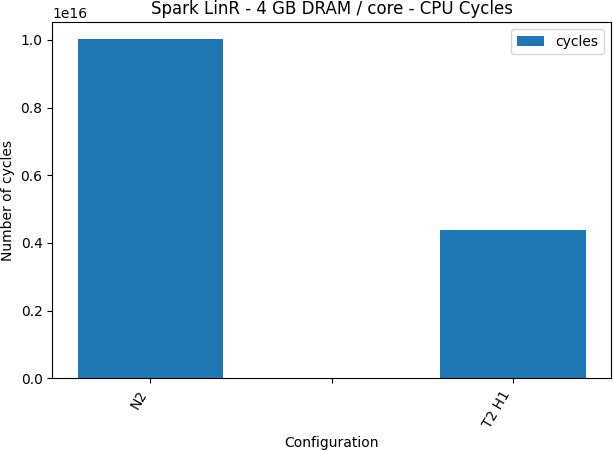}
    \caption{Native and TeraHeap Spark CPU cycles under 4 GB DRAM per core running Linear Regression.}
                \label{fig:linr32_cycles}
\end{figure}

\begin{figure}[thbp]
        \centering
        \includegraphics[width=\linewidth]{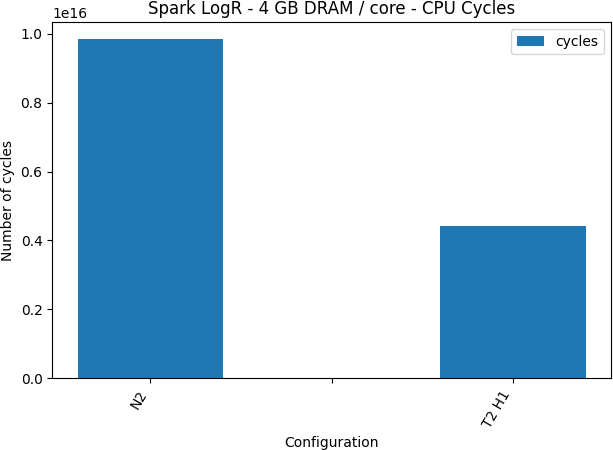}
    \caption{Native and TeraHeap Spark CPU cycles under 4 GB DRAM per core running Logistic Regression.}
                \label{fig:logr32_cycles}

        \includegraphics[width=\linewidth]{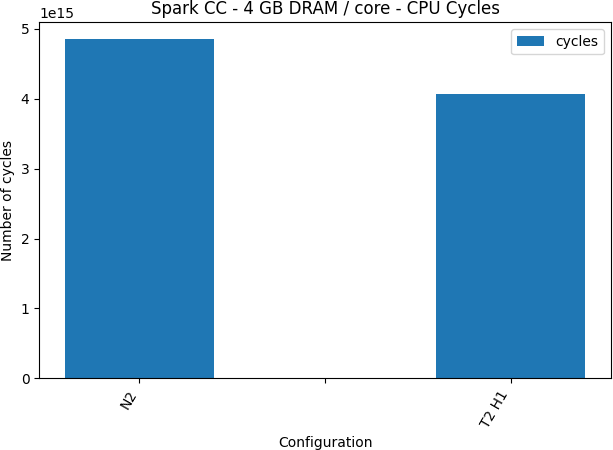}
    \caption{Native and TeraHeap Spark CPU cycles under 4 GB DRAM per core running Connected Component.}
                \label{fig:cc32_cycles}
\end{figure}

\begin{figure}[thbp]
        \centering
        \includegraphics[width=\linewidth]{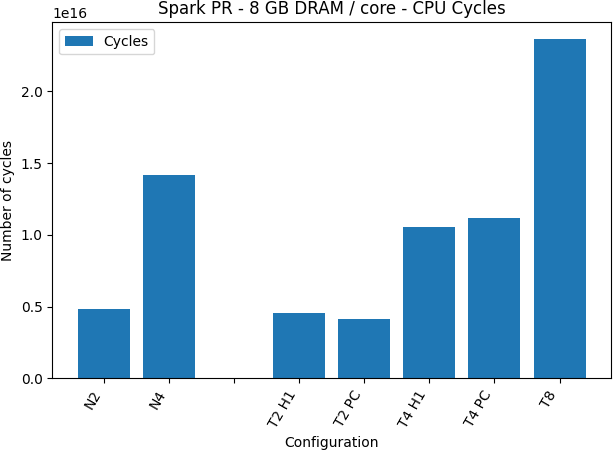}
    \caption{Native and TeraHeap Spark CPU cycles under 8 GB DRAM per core running Page Rank.}
\label{fig:pr64_cycles}
        \includegraphics[width=\linewidth]{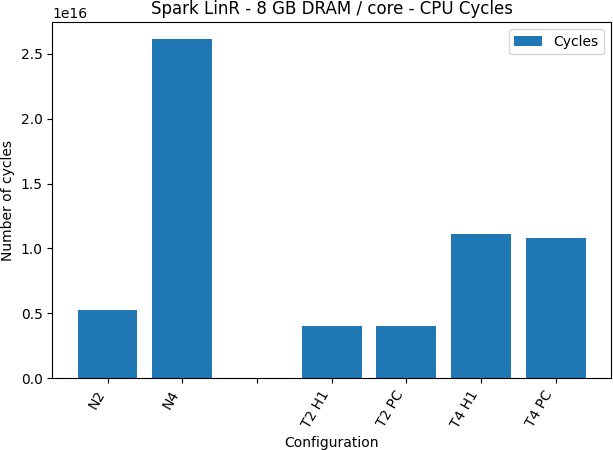}
    \caption{Native and TeraHeap Spark CPU cycles under 8 GB DRAM per core running Linear Regression.}
                \label{fig:linr64_cycles}
\end{figure}

\begin{figure}[thbp]
        \centering
        \includegraphics[width=\linewidth]{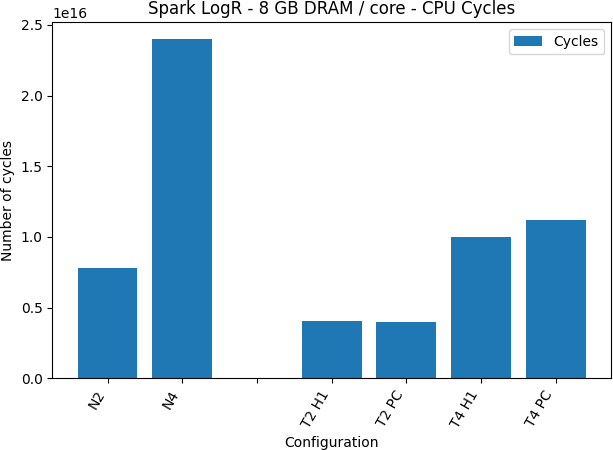}
    \caption{Native and TeraHeap Spark CPU cycles under 8 GB DRAM per core running Logistic Regression.}
                \label{fig:logr64_cycles}

        \includegraphics[width=\linewidth]{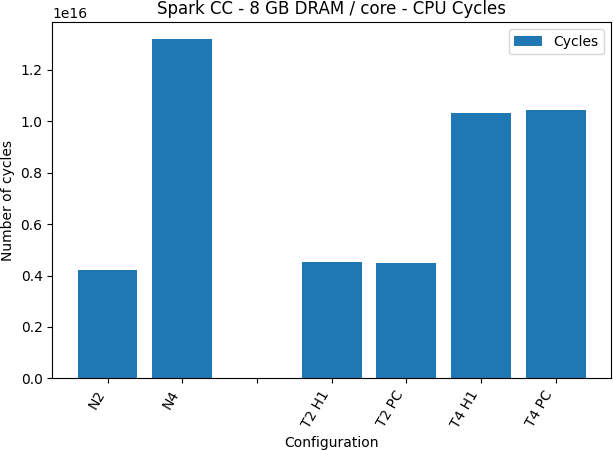}
    \caption{Native and TeraHeap Spark CPU cycles under 8 GB DRAM per core running Connected Component.}
                \label{fig:cc64_cycles}
\end{figure}

\begin{figure}[thbp]
        \centering
        \includegraphics[width=\linewidth]{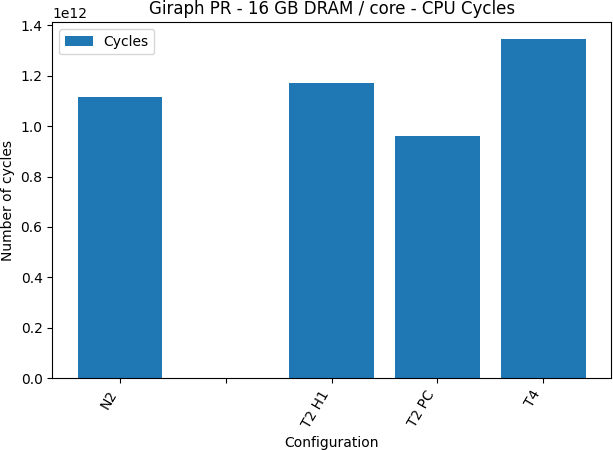}
    \caption{Native and TeraHeap Giraph CPU cycles under 16 GB DRAM per core running Page Rank.}
                \label{fig:g_pr128_cycles}

        \includegraphics[width=\linewidth]{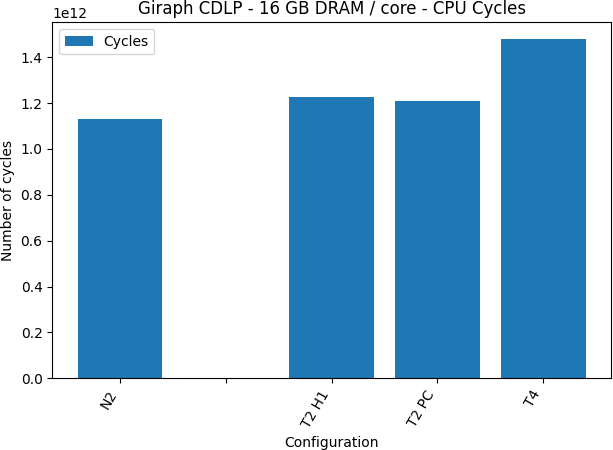}
    \caption{Native and TeraHeap Giraph CPU cycles under 16 GB DRAM per core running Community Detection Label Propagation.}
                \label{fig:g_cdlp128_cycles}
\end{figure}

\begin{figure}[thbp]
        \centering
        \includegraphics[width=\linewidth]{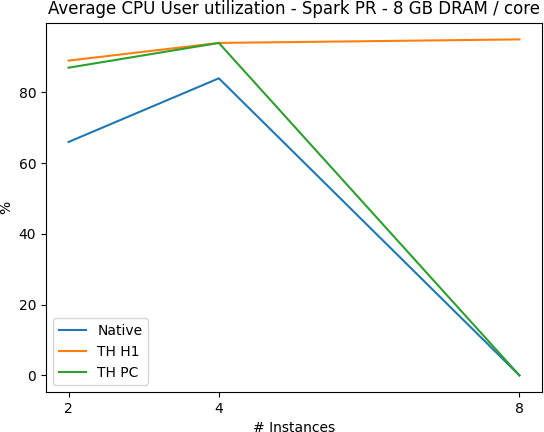}
    \caption{Native and TeraHeap Spark average user CPU utilization
        as the number of instances increases under 8 GB DRAM per core running Page Rank.}
                \label{fig:pr_64_usr}

        \includegraphics[width=\linewidth]{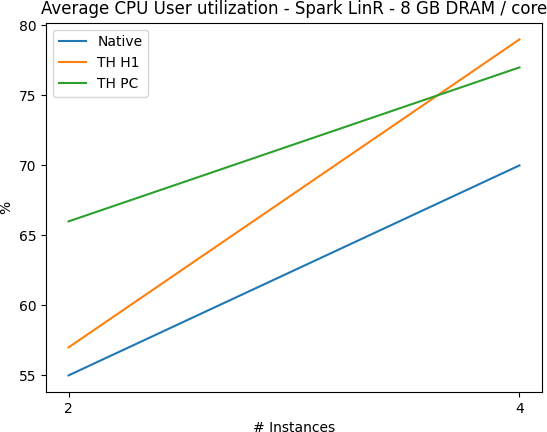}
    \caption{Native and TeraHeap Spark average user CPU utilization
        as the number of instances increases under 8 GB DRAM per core running Linear Regression.}
                \label{fig:linr_64_usr}
\end{figure}

\begin{figure}[thbp]
        \centering
   \includegraphics[width=\linewidth]{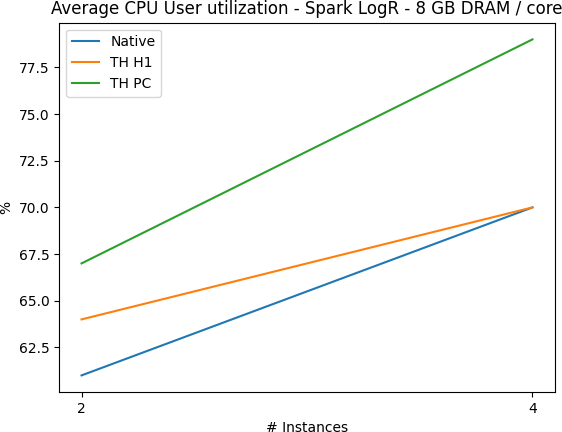}
    \caption{Native and TeraHeap Spark average user CPU utilization
        as the number of instances increases under 8 GB DRAM per core running Logistic Regression.}
           \label{fig:logr_64_usr}
	\includegraphics[width=\linewidth]{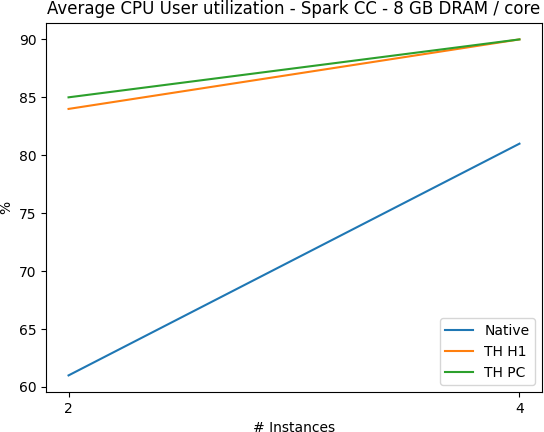}
    \caption{Native and TeraHeap Spark average user CPU utilization
        as the number of instances increases under 8 GB DRAM per core running Connected Component.}
        \label{fig:cc_64_usr}
\end{figure}

\begin{figure}[thbp]
	\centering
        \includegraphics[width=\linewidth]{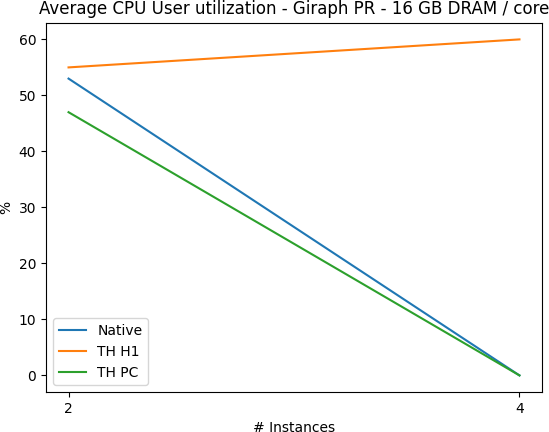}
    \caption{Native and TeraHeap Giraph average user CPU utilization
        as the number of instances increases under 16 GB DRAM per core running Page Rank.}
        \label{fig:g_pr128_usr}
        \includegraphics[width=\linewidth]{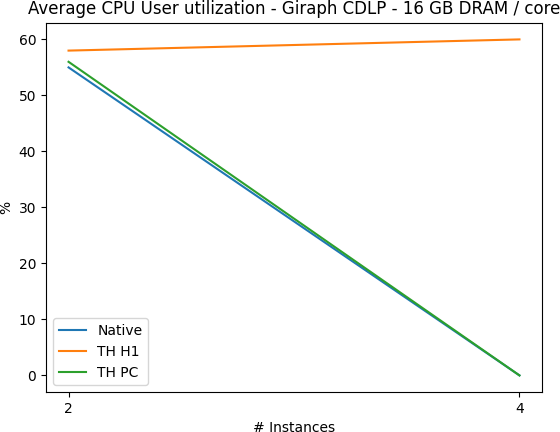}
    \caption{Native and TeraHeap Giraph average user CPU utilization
        as the number of instances increases under 16 GB DRAM per core running Page Rank.}
        \label{fig:g_cdlp128_usr}
\end{figure}

The main goal for co-locating tasks is to increase the CPU utilization and achieve better
throughput. In this section, we examine if the CPU utilization translates to better application throughput.
CPU utilization is split to 2 parts. 
User utilization includes all CPU cycles that were executed in user-space threads.
It includes GC cycles, S/D cycles and cycles for mutator tasks except I/O.
System utilization includes all CPU cycles that were executed in kernel-space threads.
This includes I/O carried out by GC (TeraHeap) and mutator I/O.
Therefore, we have to focus to User utilization, which includes the effective CPU cycles executed by the application.
We look at the CPU cycles performed by each configuration and compare it with user and total CPU utilization and then come to our conlusion.
CPU cycles are calculated using the formula (total number of cores * cpu frequency * execution time of slowest instance * cpu utlization achieved by all instances).

In the figures \ref{fig:pr32_cycles}, \ref{fig:linr32_cycles}, \ref{fig:logr32_cycles} and \ref{fig:cc32_cycles}, we look at the CPU cycles under 4 GB memory per core for Spark. We see that TH Spark executes in less CPU cycles (56\% for LinR,55\% for LogR and 16\% for CC) except for PageRank, where Native executes in less cycles by 11\%. In the same time, it has increased CPU utilization compared to Native Spark by 40, 4, 13 and 7 \% accordingly.
This means that reducing GC and S/D leads to more effective CPU utilization for all workloads except PageRank. For PageRank, TH executes in more cycles thus we cannot be sure about the benefit.
In the figures \ref{fig:pr64_cycles}, \ref{fig:linr64_cycles}, \ref{fig:logr64_cycles} and \ref{fig:cc64_cycles}, we look at the CPU cycles under 8 GB memory per core for Spark. For PR, TH Spark executes in less CPU cycles (6\% for T2 H1, 14\% for T2 PC, 25\% for T4 H1, 21\% for T4 PC).
For LinR, TH Spark executes in less CPU cycles (23\% for T2 H1, 24\% for T2 PC, 58\% for T4 H1, 59\% for T4 PC).
For LogR, TH Spark executes in less CPU cycles (48\% for T2 H1, 49\% for T2 PC, 58\% for T4 H1, 53\% for T4 PC).
For PR, TH Spark executes in less CPU cycles for 4 co-located instances (22\% for T4 H1, 21\% for T4 PC), while Native Spark executes in less
CPU cycles for 2 co-located instances (7\% against both T2 H1 and T2 PC).
This means that reducing GC and S/D leads to more effective CPU utilization for all workloads for 2 and 4 co-located instances except for CC with 2 co-located instances.
In the figures \ref{fig:g_pr128_cycles} and \ref{fig:g_cdlp128_cycles}, we look at the CPU cycles under 16 GB memory per core for Giraph. We see that TH Giraph executes in more CPU cycles except for T2 PC in PR with speedup in cycles by 14\%.
This means that reducing GC for Giraph does not necessarily lead to more effective CPU utilization.

If we look at the figures \ref{fig:pr_64_usr}, \ref{fig:linr_64_usr}, \ref{fig:logr_64_usr}, \ref{fig:cc_64_usr}, \ref{fig:g_pr128_usr} and \ref{fig:g_cdlp128_usr}, we witness User utilization for 8 GB memory per core for Spark and 16 GB memory per core for Giraph. TH has more User utilization in all scenarios.
We also include the total CPU utilization (User+System) in \ref{fig:pr_64_util}, \ref{fig:linr_64_util}, \ref{fig:logr_64_util}, \ref{fig:cc_64_util}, \ref{fig:g_pr_128_util} and \ref{fig:g_cdlp_128_util}.
For 4 GB memory per core in Spark and for 8 GB memory per core for Giraph, we do not include user utilization as the number of instances increases, since we cannot run more than 2 instances, especially for Giraph, where Native is not able to run at all. For Spark, TH increases User and and total CPU utilization accordingly to 8 and 16 GB memory per core.
By combining cycles and user utilization, we come to the conclusion that, since TH has increased User utilization in all scenarions, in the ones, where it executes in less CPU cycles it has more effective CPU utilization. That is because of reduced GC and S/D. In the scenarios where it executes in more cycles we cannot say for sure, despite TH having more average throughput. However, for Giraph, we see that decreasing GC and S/D, allows us to run more instances in the server, because TH needs less memory per instance. In terms of choosing what is best for TH, H1 or PC, we see from the CPU cycles that for Spark there are no clear benefits for any side. For Giraph, in \ref{fig:g_pr128_cycles} we see that with 4 instances PC executes in less cycles, but the execution time is the same and CPU utilization is more for TH H1 so the benefit is not clear.
To conclude for Native Spark and Giraph, we see in most scenarios that the increment in CPU utilization is not useful work, but more GC and S/D since the memory for each instance decreases as the number of co-located instances increase.

\subsection{What
happens with monetary cost across different cloud platforms?}

\begin{table}[t!]
  \centering
  \caption{Hourly costs for EC2, GCP and AZ=Azure Cloud}
  \label{tab:cost_table}
  \begin{tabular}{|c|c|c|c|}
    \hline
	  \textbf{Provider} & \textbf{DRAM (GB)} & \textbf{Cores \#} & \textbf{Hourly cost (\$)} \\
    \hline
	  EC2 & 128 & 8 & 0.67  \\
	  EC2 & 64 & 8 & 0.4 \\
	  EC2 & 32 & 8 & 0.27 \\
	  GCP & 128 & 8 & -- \\
	  GCP & 64 & 8 & 0.36 \\
	  GCP & 32 & 8 & 0.27 \\
	  AZ & 128 & 8 & 1.05 \\
	  AZ & 64 & 8 & 0.48 \\
	  AZ & 32 & 8 & 0.33 \\ 
	  \hline
  \end{tabular}
\end{table}

Tables \ref{tab:cost_table} shows hourly cost for each machine configuration in 
Amazon Web Services Cloud (EC2), GCP (Google Cloud Platform) and Microsoft Azure costs. 
We witness that Amazon and Google providers offer a similar cost for identical machines to our server.
Azure is more expensive, especially for the 16 GB memory per core machine, which is 36\% more expensive than EC2's.
Google does not offer a 16 GB memory per core machine.
Taking into account that we have an hourly cost and that we have an estimation,
reducing GC and S/D achieves benefits of up to 50\% for running co-located workloads in these clouds. The calculations are very simple so we skip them. We multiply hourly cost by number of hours needed to execute each experiment until all instances finish execution. The conclusion is that reducing GC and S/D makes a huge difference in the execution time and therefore running with TeraHeap decreases the hours needed to rent the machines. This leads to not wasting money on overheads, but using it to do actual work.

%% file: future_work.tex
\section{Future Work}

While this analysis shows promising results and provides a methodology for understanding
throughput for big data analytics workloads on Spark and Giraph
clusters, there are several avenues for future work to use it on and
improve performance and scalability.

Firstly, one potential direction for future work is to investigate the
use of other types of storage mediums such as the hybrid NVM. This
medium could improve the performance of Big data analytics further by
combining the advantages of memory and storage.

Secondly, another area for future work is to develop techniques for
dynamically adjusting the heap offloading decisions based on workload
characteristics and resource availability. For example, the offloading
decision can be based on the size of the input data or the
availability of DRAM capacity in the cluster. Such techniques can help
maximize the performance gains achieved by offloading while minimizing
the cost of offloading.

Thirdly, an interesting direction for future work is to explore the
use of heap offloading in environments where Spark-Giraph clusters are
deployed across multiple machines using RDMA to achieve communication
between the different machines. This can help utilize the DRAM, CPU
and storage availability in more than one machine and provide a more
cost-effective solution for big data processing.

Finally, investigating the power consumption of our experiments would be
very interesting, because we would examine the trade-offs between
better performance and higher resource utilization with the cost in power.

%% file: conclusion.tex
\section{Conclusions}

In this thesis, we conducted an analysis of throughput for managed big data analytics frameworks
using Apache Spark and Giraph under workload co-location. We investigated, if reducing GC and S/D for managed big data frameworks improves application throughput by using an open-source system TeraHeap. We conducted our experiments under 3 different memory-per-core
scenarios, 4, 8 and 16 GB / core, in order to see if increasing memory capacity helps increasing server throughput. 4 GB / core is the current trend and 8 and 16 GB / core are possible future trends.
For simplicity, we divided total DRAM capacity to 2,4 and 8 even memory budgets. We used each budget to run each instance isolated with Native Spark and Giraph and Spark and Giraph using TH to study the execution breakdown.
Then we run experiments with 2,4 and 8 co-located instances using the above budgets for each instance. We ran 4 Spark workloads (PR, LinR, LogR and CC) in the 4 and 8 GB / core scenario and 2 Giraph workloads (PR, CDLP) in the 8 and 16 GB / core scenario. We ran Giraph under 16 GB / core, because it is more memory intensive than Spark. We reported interference with single instance, execution breakdown (GC, S/D, I/O), user and CPU utilization, CPU cycles and average throughput. We also included a cost estimation of the experiments in several public clusters to show that decreasing GC and S/D helps utilizating monetary budgets for renting servers more effectively.

Our experimental results showed that reducing GC and S/D for Spark reduces execution time and increases the effective CPU utilization by the applications threads, where in Giraph that assumption is not confirmed. Furthermore, decreasing GC and S/D allows a higher number of co-located instances to be executed in the server, because of lower memory per instance needs. Overall, our analysis showed that high CPU utilization does not always mean that useful work is done by the CPU. Specificaly for managed
big data frameworks like Spark and Giraph a lot of CPU cycles are wasted on GC and S/D and even increasing H1 by increasing memory-per-core does not guarantee optimal execution.